\documentclass[a4paper,11pt]{article}

\usepackage{geometry}									% [showframe]
\usepackage[english]{babel}		% language
\usepackage[utf8]{inputenc}		% input text encoding
\usepackage[T1]{fontenc}		% output encoding

\usepackage{lmodern}			% Latin Modern

\usepackage{amssymb,amsfonts,amsmath,amsthm}
\usepackage{booktabs}
\usepackage{hyphenat}
\usepackage{booktabs}
\usepackage{bbm}
\usepackage{multirow}
\usepackage{graphicx}
\usepackage{algpseudocode}
\usepackage{algorithm}

\makeatletter
\let\oldabs\abs
\def\abs{\@ifstar{\oldabs}{\oldabs*}}
\let\oldnorm\norm
\def\norm{\@ifstar{\oldnorm}{\oldnorm*}}
\makeatother

\newcommand{\innerprod}[2]{\left\langle #1, #2 \right\rangle}
\renewcommand{\vec}[1]{\boldsymbol{#1}}

\newcommand{\eg}{e.\,g.}
\newcommand{\ie}{i.\,e.}
\newcommand{\wrt}{w.\,r.\,t.}

\usepackage{authblk}						% [auth-sc,affil-it], to be loaded after titling
\title{Surrogate modeling with functional nonlinear autoregressive models $\mathcal{F}$-NARX}
\author[1]{Styfen Schär\thanks{styfen.schaer@ibk.baug.ethz.ch}}
\author[1]{Stefano Marelli\thanks{marelli@ibk.baug.ethz.ch}}
\author[1]{Bruno Sudret\thanks{sudret@ethz.ch}}
\affil[1]{Chair of Risk, Safety and Uncertainty Quantification, ETH Z\"{u}rich, Switzerland}

\date{\today}

\usepackage{microtype}			% typographical refinements

\usepackage{indentfirst}		% indents first paragraph in new section

\usepackage{caption}
\usepackage{subcaption}

\usepackage{natbib}
\usepackage{hyperref}
\hypersetup{
pdftitle = {Surrogate modeling with functional nonlinear autoregressive models $\mathcal{F}$-NARX},
pdfauthor = {Styfen Sch\"ar, Stefano Marelli, Bruno Sudret},
pdfsubject = {},
pdfkeywords = {NARX, Dynamical systems, Surrogate modeling, Autoregressive modeling, Principal component analysis, Least angle regression},
colorlinks = false,			% false: boxed links; true: colored links
}

\usepackage[capitalize]{cleveref}							% Eq., Fig., Table, Section, has to be loaded after hyperref
\creflabelformat{equation}{#2(#1)#3}						% hyperlinks covers the parenthesis around equation
\crefrangelabelformat{equation}{#3(#1)#4 to #5(#2)#6}		% multiple references
\labelcrefformat{subequation}{#2(#1)#3}						% the same for subequations
\labelcrefrangeformat{subequation}{#3(#1)#4 to #5(#2)#6}	% mutiple references

\graphicspath{{.},{./figures}} 

\begin{document}

\maketitle

\begin{abstract}
We propose a novel functional approach to surrogate modeling of dynamical systems with exogenous inputs. This approach, named Functional Nonlinear AutoRegressive with eXogenous inputs ($\mathcal{F}$-NARX), approximates the system response based on temporal features of the exogenous inputs and the system response. This marks a major step away from the discrete-time-centric approach of classical NARX models, which determines the relationship between selected time steps of the input/output time series. By modeling the system in a time-feature space, $\mathcal{F}$-NARX takes advantage of the temporal smoothness of the process being modeled, providing more stable predictions and reducing the dependence of model performance on the discretization of the time axis.

In this work, we introduce an $\mathcal{F}$-NARX implementation based on principal component analysis and polynomial regression. To further improve prediction accuracy, we also introduce a modified hybrid least angle regression approach to identify a sparse model structure and minimize the expected forecast error, rather than the one-step-ahead prediction error.

We investigate the behavior and capabilities of our $\mathcal{F}$-NARX implementation on two case studies: an eight-story building under wind loading and a three-story steel frame under seismic loading. Our results demonstrate that $\mathcal{F}$-NARX has several favorable properties that make it well-suited to surrogate modeling applications.
\end{abstract}

\section{Introduction}\label{sec:introduction}
Many real-world problems are inherently time-dependent, with systems continuously evolving under the influence of external factors and excitations. 
Accurate modeling of these dynamical systems is essential in numerous engineering disciplines \citep{Cheng_2025}, be it for system control \citep{levin_1996, Hu_2024}, maintenance planning \citep{langeron_2021, Theissler_2021, Samsuri_2023, Dehghan_2024}, fault detection and diagnosis \citep{mattson_2006, Gao_2016, Henrique_2021}, design assessment and optimization \citep{Yu_2023, deshmukh_2017} or uncertainty quantification \citep{mai_2016, bhattacharyya_2020, wu_2025} and reliability analysis \citep{garg_2022, li_2022, Zhou_2023, Zhang2_2024}.

Although these applications are different in nature, they are similar in that they often require to learn the dependence between the time-varying external factors or actions and the corresponding system response. 
This dependence is often modeled by so-called autoregressive with exogenous inputs (ARX) models \citep{billings_2013}.
By using the exogenous inputs in conjunction with their own past predictions, these models are powerful predictors for dynamical systems, especially when used in their nonlinear variant, known as NARX. 
NARX models have been successfully applied to many engineering systems structures and components, such as mooring lines \citep{Yetkin_2017, Zhang_2024}, gas or wind turbines \citep{chiras_2001, schlechtingen_2011}, multi-story buildings \citep{spiridonakos_2015, Li_2021b}, marine structures \citep{Kim_2015, Poursorkhabi_2023} or geotechnical systems \citep{Wunsch_2018, ma_2020, Dassanayake_2023}.

This success is partially due to their versatility, which allows one to combine them with powerful techniques from the fields of surrogate modeling and machine learning, such as Gaussian process modeling \citep{murray_1999, kocijan_2012, koziel_2014, worden_2018}, support vector regression \citep{Gonzalo_2012, rankovic_2014, Zhang_2017} or neural networks \citep{Siegelmann_1997, li_2021, song_2022}. 
In a \emph{linear-in-the-parameters} setting \citep{billings_2013} they can also be combined with a wide array of basis functions, such as polynomials \citep{Aguirre_1993}, wavelets \citep{Coca_2001} or radial basis functions \citep{chen_1990}.
They have also been successfully deployed in applications with categorical outcomes by combining them with logistic regression \citep{Ayala_2017, Rocha_2021}.

Fitting a predictive model to existing data is generally a difficult task, and since NARX models rely on the time discretization of the problem at hand, they pose one additional key challenge: the selection of lags to be considered. 
In other words, how to select which time instants of the exogenous and autoregressive inputs are to be considered in the model?
This selection has to be made from a possibly large set of candidate lags, for example, if the time-discretization of the problem is small or the model's memory long. 
While no universal solution to this problem is currently available from the relevant literature, several ways to directly select important lags, or at least decimate the candidate lags, have been proposed. Examples include trial-and-error-based approaches \citep{chen_2011, Schaers_2024, Awtoniuk_2019} or correlation-based approaches \citep{wei_2008, cheng_2011}.
An approach named ``clustering'' has been proposed by \citet{Aguirre_1995b, Aguirre_1998} for the specific case of polynomial NARX models.
For the special class of linear-in-the-parameters problems \citep{billings_2013}, dedicated sparse solvers have also been employed to better handle a large number of regressors \citep{Spinelli_2006, billings_2013, Falsone_2014, zhang_2015, Guo_2015, mai_2016, Bianchi_2017, Li_2021b}. 

NARX models are most often trained by minimizing the one-step-ahead prediction (OSA) error, as this is very fast and sometimes even a closed-form solution exists. 
However, minimization of OSA errors can lead to poor results when the trained model is used in a forecast setup \citep{Piroddi_2008}.
In extreme cases, the model predictions can even diverge over time \citep{Piroddi_2008, Farina_2009, Yu_2023}. 
Therefore, dedicated solvers have been developed to minimize the forecast error. 
For instance, \citet{Piroddi_2003, Piroddi_2008, Piroddi_2010} present pruning-based methods, while \citet{Farina_2009, Farina_2010} introduced gradient-based algorithms. 
The main disadvantage of these approaches is the heavily increased computational costs compared to the minimization of the OSA error. 

Several works \citep{Aguirre_1994, Billings_1995, Spinelli_2006} argue that the negative consequences of an OSA error minimization are more likely to happen if the sampling frequency of the problem is high. 
The reason lies in the high temporal correlation between adjacent time steps, which in turn can lead to the over-reliance on the first autoregressive lag, resulting in drastic losses in terms of the final model performance \citep{Piroddi_2003}. 

Interestingly, all these limitations are related to the discretization of the time axis, in conjunction with the exclusive focus of classical ARX modeling on individual discrete time step values, a characteristic we refer to as a \emph{discrete-time} view on ARX modeling. 
Although many methods have been developed to mitigate these problems, they remain fully committed to this discrete-time view.
This raises the question of whether alternative views can offer advantages over the discrete-time one.

In this work, we revisit the fundamentals of autoregressive problems with exogenous inputs and propose a novel \emph{continuous functional} view on dynamical systems.
This continuous functional approach leverages the temporal correlation and smoothness exhibited by many time-dependent processes, especially in physical systems or structural responses.
Due to its underlying principles, we refer to this approach as \emph{functional nonlinear autoregressive with exogenous inputs} ($\mathcal{F}$-NARX) modeling.

$\mathcal{F}$-NARX modeling addresses the problem of lag selection, mitigates over-reliance, and thus improves forecast stability and accuracy.
Its use of temporal features of the input and output signals allows the problem to still be posed as a linear regression problem, and therefore $\mathcal{F}$-NARX remains compatible with most algorithms developed for classical NARX modeling.
Moreover, temporal features can lead to sparser representations of both input and output signals and promote independence between regressors, which can be exploited through compressed sensing algorithms \citep{Eldar_2012} to create compact models with high expressiveness.

While $\mathcal{F}$-NARX is applicable to a broad range of popular tasks, including system identification and time-series forecasting common in machine learning applications, this work focuses on its applicability to surrogate modeling. 
In surrogate modeling, the surrogate is typically trained on multiple evaluations of a physics-based model at different input configurations, with the goal of predicting the whole response trajectory for new unseen inputs.
This contrasts with system identification and time-series forecasting, where the model is often trained on a single observed sequence and used to only forecast its future evolution. 
Surrogate modeling thus poses a particularly challenging problem, as it requires significant extrapolation capability and stability, thus pushing the limits of model generalization, making it an ideal test case to evaluate the performance of $\mathcal{F}$-NARX.
To this end, the paper is structured as follows: Sections~\ref{sec:methodology} and \ref{sec:sparse_fnarx} provide a detailed description of the $\mathcal{F}$-NARX approach and its sparse formulation.
This includes a brief recap on the basics of NARX modeling, followed by its extension to $\mathcal{F}$-NARX, as well as a concrete implementation of an $\mathcal{F}$-NARX model using a combination of principal-component analysis and a polynomial regression model.
In Section~\ref{sec:applications}, we present two surrogate modeling case studies to investigate the behavior and properties of the presented $\mathcal{F}$-NARX implementation and demonstrate its performance on a complex dynamical system. 
A discussion on the $\mathcal{F}$-NARX approach and the presented results, as well as concluding remarks, are given in Section~\ref{sec:discussion_and_conclusion}.

\section{Methodology}\label{sec:methodology}

\subsection{Autoregressive modeling with exogenous inputs}\label{sec:arx_modeling}
Consider a deterministic dynamical system $\mathcal{M}$ evolving along the time axis $\mathcal{T}$. 
The system is excited by a (possibly high-dimensional) time-varying exogenous input $\vec{x}(t) \in \mathbb{R}^{M}$. Given a vector of initial conditions $\vec{\beta}$, the system response $y(t) \in \mathbb{R}$ at any time instant $t \in \mathcal{T}$ is denoted as:
\begin{equation}
	y(t) = \mathcal{M}(\vec{x}(\mathcal{T} \le t), \vec{\beta}).
\end{equation}
Here, the notation $\bullet(\mathcal{T} \le t)$ indicates that the system response at time $t$ depends on the excitation up to and including time $t$. 
To simplify the notation, we will subsequently omit $\vec{\beta}$ unless it is strictly necessary.

Suppose we want to construct a dynamic surrogate model $\widehat{\mathcal{M}}$ that approximates the response of the system $\mathcal{M}$ such that:
\begin{equation}\label{eq:dynamic surrogate}
	\widehat{y}(t) = \widehat{\mathcal{M}}(\vec{x}(\mathcal{T} \le t), \vec{\beta}) \approx \mathcal{M}(\vec{x}(\mathcal{T} \le t), \vec{\beta}).
\end{equation}
In this work, we focus in particular on dynamic surrogates based on nonlinear autoregressive models with exogenous inputs (NARX). 
At the core of a NARX model lies the idea that the system's dynamics can be captured at a set of discrete time steps $\{0, \delta t, \dots, (N-1)\delta t\}$ and that the system response at a time step in the near future can be predicted as a function of past and current exogenous inputs, as well as past outputs:
\begin{equation}\label{eq:NARX next timestep}
	\widehat{y}(t+\delta t) = \widehat{\mathcal{M}}(\vec{x}(\mathcal{T} \le t+\delta t), y(\mathcal{T} < t+\delta t); \vec{c}) + \varepsilon(t),
\end{equation}
where $\varepsilon(t) \sim \mathcal{N}(0, \sigma_{\varepsilon}(t))$ is a residual term with zero mean and standard deviation $\sigma_{\varepsilon}(t)$. 
Let us assume the mapping function $\widehat{\mathcal{M}}$ is parametric, and that it can be characterized by a finite set of parameters $\vec{c}$. Then the process of training (or fitting) a NARX model consists in estimating the values of $\vec{c}$ from a set of system input/output discretized trajectories $\left( \vec{x}^{(i)}, \vec{y}^{(i)} \right)$. We refer to each such pair as a \emph{realization} or \emph{observation} of the system, while the full set of realizations is called the \emph{experimental design} (ED):
\begin{equation}\label{eq:experimental_design}
	\mathcal{D} = \left\{ \left( \vec{x}^{(i)}, \vec{y}^{(i)} \right) , \vec{x}^{(i)} \in \mathbb{R}^{N \times M}, \vec{y}^{(i)} = \mathcal{M}(\vec{x}^{(i)}) \in \mathbb{R}^{N}, i=1, \dots, N_\text{ED} \right\}.
\end{equation}
The number of realizations in the experimental design is usually relatively small ($\mathcal{O}(10^{1-2})$), since the acquisition of even a single observation may require an expensive experiment or simulation. 

We can make the notation more explicit by rewriting Eq.~\eqref{eq:NARX next timestep} as:
\begin{equation}\label{eq:discrete_prediction}
	\widehat{y}(t+\delta t) = \widehat{\mathcal{M}}(\vec{\varphi}(t+\delta t); \vec{c}),
\end{equation}
where the vector $\vec{\varphi}(t) \in \mathbb{R}^{n}$ reads:
\begin{equation}\label{eq:varphi}
	\begin{split}
		\vec{\varphi}(t) = \{
		&y(t - \delta t), y(t - 2\delta t), \dots, y(t - n_y\delta t), \\
		&x_1(t), x_1(t - \delta t), \dots, x_1(t - n_{x_1}\delta t), \\ 
		&\dots, \\
		&x_M(t), x_M(t - \delta t), \dots, x_M(t - n_{x_M}\delta t)\}.
	\end{split}
\end{equation}
We refer to the delayed values $y(t-(k+1)\delta t)$ as the autoregressive lags and to $x_i(t-k\delta t)$ as the exogenous input lags, where $k$ is a non-negative integer value to preserve the causality of the original system. 
The integers $\{n_y, n_{x_1}, \dots, n_{x_M}\}$ are typically referred to as the \emph{model orders}.

By stacking the vectors $\vec{\varphi}(t)$ for all time steps, we obtain the so-called \emph{design matrix} $\vec{\Phi} \in \mathbb{R}^{\widetilde{N} \times n}$, where $n = n_y + n_{x_1} + \dots + n_{x_M}$ and where $\widetilde{N} \le N$ due to the use of lagged values.
Similarly, we stack the values $y(t)$ to obtain the corresponding output vector $\vec{y} \in \mathbb{R}^{\widetilde{N}}$:
\begin{equation}\label{eq:Phi_matrix}
	\vec{\Phi} = \begin{pmatrix}
		\vec{\varphi}(t_0) \\
		\vec{\varphi}(t_0+\delta t) \\
		\vdots \\
		\vec{\varphi}((N-1)\delta t)
	\end{pmatrix},
	\quad
	\vec{y} = \begin{pmatrix}
		y(t_0) \\
		y(t_0+\delta t) \\
		\vdots \\
		y((N-1)\delta t)
	\end{pmatrix},
\end{equation}
with $t_0 = \max(n_y, n_{x_1}, \dots, n_{x_M}) \delta t$.

Note that despite considering time-dependent data, the sample pairs $\{ \vec{\varphi}(t), y(t) \}$ no longer need to follow a temporal ordering.
Consequently, the matrices $\vec{\Phi}^{(i)}$ and vectors $\vec{y}^{(i)}$, with $i=1, \dots, N_\text{ED}$, from different realizations within the experimental design can be concatenated to form a larger matrix $\vec{\Phi}_\text{ED}$ and vector $\vec{y}_\text{ED}$:
\begin{equation}\label{eq:large_Phi_matrix}
	\vec{\Phi}_\text{ED} = \begin{pmatrix}
		\vec{\Phi}^{(1)} \\
		\vdots \\
		\vec{\Phi}^{(N_\text{ED})}
	\end{pmatrix}, 
	\quad
	\vec{y}_\text{ED} = \begin{pmatrix}
		\vec{y}^{(1)} \\ 
		\vdots \\
		\vec{y}^{(N_\text{ED})} 
	\end{pmatrix}.
\end{equation}

We now recall from Eq.~\eqref{eq:NARX next timestep}, that the objective is to develop a predictive model $\widehat{\mathcal{M}}$, characterized by its parameters $\vec{c}$, using only a limited set of observations $\mathcal{D}$.
The estimation of these model parameters can then be performed by minimizing a suitable loss function $\mathcal{L}$ that represents the discrepancy between the experimental design observations and the model predictions:
\begin{equation}\label{eq:loss_minimization}
	\hat{\vec{c}} = \mathop{\arg\min}_{\vec{c}} \mathcal{L}\left(\vec{y}_\text{ED}, \widehat{\mathcal{M}}(\vec{\Phi}_\text{ED}; \vec{c})\right).
\end{equation}
Common choices for $\widehat{\mathcal{M}}$ include polynomials, for which $\vec{c}$ represents the coefficient vector \citep{Aguirre_1993},  neural networks, with $\vec{c}$ denoting the corresponding weights and biases \citep{Siegelmann_1997}, among others.
A class of models frequently adopted in the literature is that of linear-in-the-parameters models \citep{billings_2013}. This class is particularly attractive, because it allows the optimization problem in Eq.~\eqref{eq:loss_minimization} to be solved using least-squares minimization:
\begin{equation}\label{eq:least_squares}
	\hat{\vec{c}} = \mathop{\arg\min}_{\vec{c}} \| \vec{y}_\text{ED} - \mathcal{G}(\vec{\Phi}_\text{ED}) \vec{c} \|^2,
\end{equation}
where $\mathcal{G}: \mathbb R^{|\vec \Phi|} \rightarrow \mathbb R^{|\vec c|}$ (with $|\cdot|$ denoting cardinality) denotes a possibly nonlinear mapping between lags and regressors, \eg\ multivariate polynomials.
By defining $\vec{\Psi}_\text{ED} = \mathcal{G}(\vec{\Phi}_\text{ED})$ the coefficient vector $\vec{c}$ can therefore be computed analytically using ordinary least squares (OLS):
\begin{equation}\label{eq:ordinary_least_squares}
		\vec{c} = \left( {\vec{\Psi}_\text{ED}}^\top {\vec{\Psi}_\text{ED}} \right)^{-1} {\vec{\Psi}_\text{ED}}^\top \vec{y}_\text{ED}.
\end{equation}
Although OLS is a viable option to solve for $\vec{c}$, more advanced regularized regression techniques are often used to promote a sparse set of coefficients, resulting in a more stable model with improved generalization capabilities. 
A prominent example is given by LASSO (least absolute shrinkage and selection operator) regression \citep{Tibshirani_1996_LASSO} which promotes sparsity introducing an $\ell^1$ penalty term with regularization parameter $\gamma$ for the estimation of the coefficient vector:
\begin{equation}\label{eq:lasso}
	\hat{\vec{c}} = \mathop{\arg\min}_{\vec{c}} \| \vec{y}_\text{ED} - \vec{\Psi}_\text{ED} \vec{c} \|^2_2 + \gamma ||\vec{c}||_1.
\end{equation}
Among the available solvers of the LASSO problem, least angle regression (LARS, \citep{Efron_2004}) has been proven particularly effective in the context of polynomial regression \citep{Blatman_2011}, and is the technique of choice in this work (see Section~\ref{sec:LARS}).

A detailed explanation and discussion on polynomial NARX models, as a concrete implementation of a linear-in-the-parameters model, and their synergy with sparse regression techniques will be provided in Sections~\ref{sec:polynomial_NARX} and \ref{sec:LARS}.

Considering again the linear-in-the-parameters model with its mapping $\mathcal{G}$ and coefficient set $\vec{c}$, we can calculate the so-called \emph{one-step-ahead} (OSA) prediction on a new input vector $\vec{\varphi}(t+\delta t)$ as:
\begin{equation}\label{eq:osa}
	\widehat{y}(t+\delta t) = \mathcal{G}(\vec{\varphi}(t+\delta t)) \vec{c}.
\end{equation}
Note, however, that in order to construct $\vec{\varphi}(t + \delta t)$, we require the true model output up to time $t$ (see Eq.~\eqref{eq:varphi}). While many applications of autoregressive modeling are focused on one-step-ahead predictions, in surrogate modeling this is typically not the case. 
With a surrogate model, we aim to approximate the true model response \emph{throughout the duration} of the external dynamic loading, \ie\ the whole trajectory. We refer to this as \emph{model forecast}, or simply as \emph{model prediction}. 
The model forecast is based on iteratively evaluating one-step-ahead predictions, and using the corresponding approximate response $\hat y(t)$ as the autoregressive input for the next step, resulting in:
\begin{equation}\label{eq:varphi_prediction}
	\begin{split}
		\widehat{\vec{\varphi}}(t+\delta t) = \{
		&\widehat{y}(t), \widehat{y}(t - \delta t), \dots, \widehat{y}(t - (n_y-1)\delta t), \\
		&x_1(t+\delta t), x_1(t), \dots, x_1(t - (n_{x_1}-1)\delta t), \\ 
		&\dots, \\
		&x_M(t+\delta t), x_M(t), \dots, x_M(t - (n_{x_M}-1)\delta t)\},
	\end{split}
\end{equation}
which contains the previous predictions $\widehat{y}(\mathcal{T} \le t)$. In order to perform this prediction, the model needs to be initialized with values that match the initial conditions of the system under investigation. This is done by setting the $n_y$ initial values of the system response to appropriate values. The surrogate prediction can be initialized to zero \citep{Worden_2012} or other sensible values \citep{Schaers_2024}, depending on the specific application.

\subsection{Limitations of classical ARX modeling}\label{sec:NARX_limitations}
If a nonlinear ARX (NARX) model is fitted as described in Eq.~\eqref{eq:loss_minimization} or \eqref{eq:least_squares}, the associated regression problem becomes intractable if the number of exogenous inputs $M$ or the model orders are large. 
This is due to the highly non-linear scaling of the number of regressors with the problem dimension, a phenomenon known as the \textit{curse of dimensionality} \citep{Verleysen_2005}.
To handle systems requiring large model orders, while still keeping the dimensionality of $\vec{\Phi}$ manageable, \citet{Schaers_2024} introduced the use of a non-evenly spaced set of lags, determined through trial and error on the available training data. 
Although it was shown to be a viable option in terms of the accuracy of the final model, trial and error can be time consuming and does not provide a good guarantee of near-optimal performance.
Alternatively, \citet{Awtoniuk_2019} proposed to select a subset of the lags following certain pre-determined patterns, for example only odd or even lags. However, this approach is limited because regularly spaced lags can cause loss of important information or aliasing effects if significant decimation is performed.

An approach limited to linear-in-the-parameters models, as described in Eq.~\eqref{eq:least_squares}, involves the use of sparse regression solvers. 
In particular, several forward selection algorithms have gained popularity \citep{billings_2013, Guo_2015, mai_2016}. Unfortunately, these sparse solvers also have practical limitations regarding the number of regressors. With a very large number of regressors, efficiency decreases, and computational costs or hardware resources can become a bottleneck.

A more specific method, called \emph{term clustering}, was developed for polynomial NARX models to discard entire clusters of lagged values belonging to irrelevant nonlinearities \citep{Aguirre_1995b, Aguirre_1998, Pulecchi_2007}. 
Selecting the correct clusters can also be particularly challenging, especially in the presence of noise.
The specificity to a certain class of NARX models may also have hindered the widespread adoption of this method.

Another difficulty arises when fitting ARX models in a regression setting when the time increment $\delta t$ is small, \eg\ due to requirements of the numerical solver of choice. 
Very small $\delta t$ can lead to numerical instability during the regression task, due to high correlation between lags. 
In \citet{Piroddi_2003} it is shown how a very high sampling rate can also lead to an overestimation of the importance of the most recent autoregressive lag. This \emph{over-reliance} can have a detrimental effect on the model forecast.
Unfortunately, data decimation to alleviate these issues is not always viable, since the data may be correctly sampled for signal reconstruction purposes and may only appear oversampled with respect to the prediction problem \citep{Piroddi_2003, lataniotis_2020}.

We observe that a common cause of all these limitations lies in the explicitly \emph{discrete-time-centric} nature of classical autoregressive modeling, which assumes a specific time axis discretization, rather than the intrinsically continuous nature of the response of dynamical systems to exogenous excitations.
In the following sections, we therefore propose and discuss a novel \emph{continuous-functional} view of the ARX methodology, which tackles autoregressive problems from a continuous perspective and discretizes the problem only in a final stage for numerical purposes.

\subsection{Moving from a discrete time to a continuous functional view of autoregressive modeling}\label{sec:feature_centric_view}

Real-world processes are inherently continuous in time, and the discretization of the time axis is merely a tool required by digital data storage or numerical simulators.
This fact is acknowledged, for example, by continuous-time system identification, which treats the dynamics of the system as a differential equation \citep{billings_2013}. 
Interestingly, NARX models also find their application in this area. 
They are used as intermediate discrete parametric models, which are then used to derive continuous-time models.
Yet again, the focus lies on the fact that the problem can be discretized in time and the NARX models do not fully exploit the continuous nature of the underlying process.

\begin{figure}[!ht]
	\centering
	\includegraphics[width=0.6\textwidth]{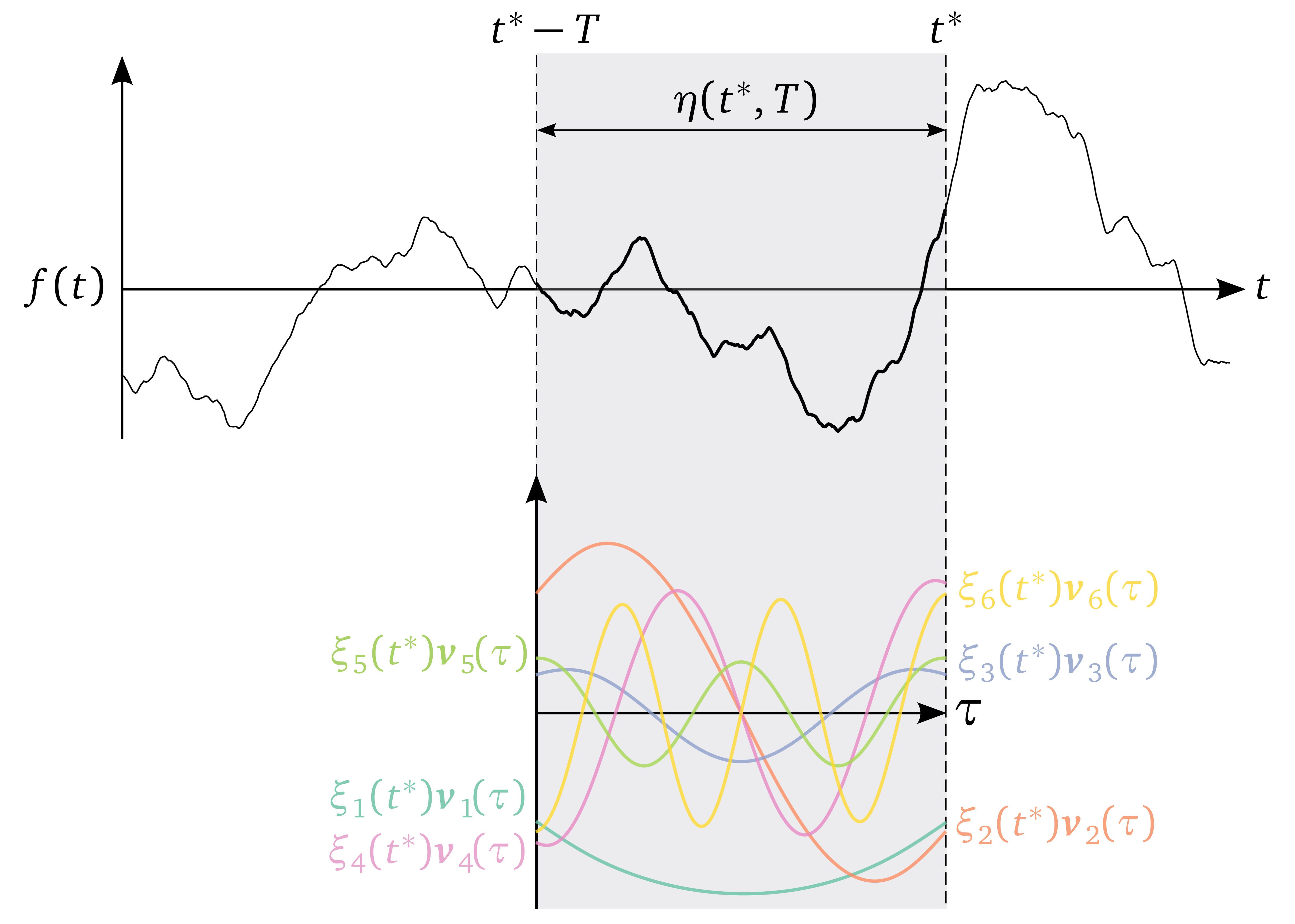}
	\caption{
		(Top) Continuous-time signal $f(t)$ indexed by the ``global'' time $t$ with highlighted time window $\eta(t^{*}, T)$ with length $T$.
		(Bottom) Temporal signal features $\vec{v}_i(\tau)$ indexed by the ``local'' time $\tau$ and scaled by the corresponding coefficients $\xi_i(t^*)$.
	}
	\label{fig:features_illustration}
\end{figure}

To devise a less discrete view of ARX modeling, let us first consider a continuous-time signal $f(t)$, as depicted in Figure~\ref{fig:features_illustration}.
We set our focus on the signal during a time window $\eta(t^{*}, T)$, indexed by a local time $\tau \in \left[ 0, T\right]$, with endpoint $t^{*}$ and finite length $T$, \ie, we define the continuous interval $\eta(t, T) = [ t - T, t ]$, such that:
\begin{equation}\label{eq:f windowed}
	f(\eta(t^{*},T),\tau) = f(t^*-T +\tau).
\end{equation}

We assume that the signal $f(\eta(t^{*}, T),\tau)$ exhibits some form of temporal regularity, and that its dynamic information can be represented by a set of continuous temporal features $\vec{v} (\tau)$, \eg\ cosine functions, wavelets, polynomials, etc., of the form:
\begin{equation}\label{eq:f windowed decomposition}
	f(\eta(t^{*},T),\tau) = \sum\limits_{i = 1}^{\infty} \innerprod{f(\eta(t^{*},T),\tau)}{v_{i}(\tau)}_\tau v_{i}(\tau) = \sum\limits_{i = 1}^{\infty} \xi_i(t^*)v_{i}(\tau),
\end{equation}
where the $\xi_i(t^{*}) \stackrel{\text{def}}{=} \innerprod{f(\eta(t^{*},T))}{v(\tau)}_{\tau} = \innerprod{f(t^{*}-T+\tau)}{v(\tau)}_{\tau}$ represent the projection of the original function in the time window $\eta (t^*,T)$ onto the $i$-th continuous temporal feature $v_i(\tau)$.
This is a mild assumption, as in practice virtually all physics-based input and output quantities in dynamic system modeling are square integrable within a bound time interval, and hence they belong to the Hilbert space $\mathcal{L}_2$.

An interesting property of the right-hand side of Eq.~\eqref{eq:f windowed decomposition} is that it explicitly decomposes the time dependence of the underlying function $f(t)$ in two different components: the scalar coefficients $\vec{\xi}(t^*)$ encoding the ``global'' time $t^*$, but not the ``local'' time $\tau$ within the window, while the continuous features $v_i$ encode the dependence on $\tau$ instead.
In other words, we decompose a moving window of duration $T$, starting from the global time $t^{*}-T$, on a set of temporal features $v_i(\tau)$. 

For illustrative purposes, the bottom plot in Figure~\ref{fig:features_illustration} illustrates such a finite set of continuous features $v_1(\tau), \dots, v_{N_F}(\tau)$ weighted by their coefficients $\xi_1(t^{*}), \dots, \xi_{N_F}(t^{*})$, such that the response of the function $f(t)$ on the highlighted time window $\eta(t^*,T)$ is decomposed as:
\begin{equation}\label{eq:projection_onto_nu}
	f(\eta(t^{*}, T),\tau)=\sum_{i=1}^{N_F}\xi_i(t^{*})v_i(\tau), ~~\tau \in [0, T].
\end{equation}

Representing a time-dependent signal through its properties within a fixed-width moving window is a well-established technique in signal processing, widely known as the \textit{sliding window} \citep{Bastiaans_1985, Badeau_2004}.
Nevertheless, to the authors' best knowledge, this is the first time this approach is combined within an autoregressive setting.
Popular choices to represent time-dependent signals in the form of Eq.~\eqref{eq:f windowed decomposition} include Fourier transforms \citep{Bracewell_1989}, and Karhunen-Lo\`eve expansions \citep{loeve_1955}. 
An advantage of structured representations like these is that real-world processes are often represented adequately even using a heavily truncated set of features.
This property is taken advantage of, for example, for dimensionality reduction \citep{Bengio_2006} or lossy signal compression \citep{Strang_1996}.
Autoregressive inputs and responses often exhibit strong temporal regularity and predictable behavior, originating from the inert physical processes that govern them, and are therefore suitable for this type of representation.
For physical systems, domain-specific knowledge may be even leveraged to produce particularly efficient representations.

In the context of autoregressive simulation, additional features may also be considered to simplify the construction of the input/output map in Eq.~\eqref{eq:NARX next timestep}. These include classical statistics in the sliding window, such as moving averages, variances, dominating frequencies, principal components, singular value decomposition eigenmodes, among others. 
The usefulness of using prior knowledge about the physics of the system in the ARX modeling process has been demonstrated in \citet{lee_2018} and \citet{Schaers_2024}.

In the following section, we will discuss how we can use features of the exogenous inputs and output to reformulate the classical discrete-time view on ARX into a continuous functional view that can mitigate or even eliminate some of its limitations.

\subsection{Functional nonlinear autoregressive with exogenous inputs modeling ($\mathcal{F}$-NARX)}
\label{sec:fnarx_modeling}

We introduce here our adaptation of NARX modeling designed to work with temporal features, rather than lags. 
We start by defining the time-dependent \emph{information vector} $\vec\zeta(t)$, that gathers all exogenous inputs together with the model response within a time window $\eta(t,T)$ as follows:
\begin{equation}\label{eq:information_vector}
	\vec{\zeta}(t) = \{ 
	x_1(\eta(t, T)),
	\dots, 
	x_M(\eta(t, T)),
	y(\eta(t-\varepsilon, T-\varepsilon))
	\},
\end{equation}
where $0 < \varepsilon \ll T$ is a small time difference used to indicate causality.

Intuitively, the window width $T$ can be considered as the \emph{memory} of the model.
Ideally, it is close to the \emph{effective memory} of the system with respect to the corresponding variable.
We define the effective memory as the look-back time such that the effect of the independent variable $\vec{x}_i$ on the dependent variable $\vec{y}$ in Eq.~\eqref{eq:NARX next timestep} becomes negligibly small.
For instance, if the system exhibits slow decay or low-frequency dynamics, a longer time window is likely required to capture its behavior. 
Since the choice of $T$ should reflect the temporal extent over which past inputs and outputs influence the system response, tools such as spectral or modal analysis can help identify dominant frequencies. A reasonable estimate for $T$ may be obtained, \eg\ by taking the period $T_0$ associated to the lowest natural frequency. Based on the presented applications and further experiments, we recommend $T \in [1.5T_0, 2T_0]$.
If $T$ is chosen too short, physically relevant dependencies that unfold over longer timescales may be missed. Selecting $T$ much larger than the effective system memory may introduce instead physically irrelevant past information, artificially inflating the problem dimension, and thus overall degrading the performance of the surrogate.

In principle, when considering complex systems with different processes and inputs, a different memory could be considered for each component of $\zeta_i(t)$ in Eq.~\eqref{eq:information_vector}, resulting in a set of memories $\vec{T} = \left\{T_1,\cdots,T_{|\vec{\zeta}|}\right\}$.
Choosing different memory lengths for different exogenous and autoregressive components may be appropriate when their effects occur at significantly different time scales. For instance, a system might respond rapidly to certain external excitations while simultaneously exhibiting a slow, resonance-like oscillatory behavior.
To simplify our notation and avoid adding complexity, we assume, without loss of generality, a uniform memory $T$ for all input and output components from this point onward.

For each of the $(M+1)$ components $\zeta_j(t)$ in Eq.~\eqref{eq:information_vector}, we now introduce a corresponding function $\mathcal{K}_j$, which extracts a set of features $\vec\xi_j(t)$: 
\begin{equation}\label{eq:xi_j}
	\vec\xi_j(t) = \mathcal{K}_j(\zeta_j(t)).
\end{equation}
No restrictions are made on the nature of the transforms $\mathcal{K}_j$, nor on the dimensionality of each feature-set $\vec\xi_j(t)$. 
Different elements of the information vector $\vec\zeta(t)$ can have different transforms, or different truncation schemes.
For example, some exogenous input $\zeta_p(t)$  may exhibit a strong cyclic behavior, and the corresponding $\vec\xi_p(t)$ may represent a subset of its Fourier coefficients, while some other input, say $\zeta_q(t)$, may instead have a smoother behavior, better described by a polynomial representation with coefficients $\vec\xi_q(t)$.

We gather all the feature sets $\vec\xi_j$ into a single feature vector $\vec\xi(t)$, with $|\vec\xi(t)| = \tilde n$:
\begin{equation}\label{eq:xi}
	\vec\xi(t) = \left\{\vec\xi_{x_1}(t), \cdots, \vec\xi_{x_M}(t), \vec\xi_{y}(t)\right\}.
\end{equation}
Note that $\tilde n$ is the total number of relevant features extracted from the $M+1$ components.
We can now adapt the classical NARX approach in Section~\ref{sec:arx_modeling}, by capitalizing on the feature vector $\vec\xi(t)$.
The goal is to approximate a near future response value $\widehat y(t+\varepsilon)\approx y(t+\varepsilon)$ as a function of $\vec\xi(t+\varepsilon)$:
\begin{equation}\label{eq:continuous_prediction}
	\widehat{y}(t+\varepsilon) = \widehat{\mathcal{M}}(\vec{\xi}(t+\varepsilon); \vec{c}).
\end{equation}

An interesting aspect of this formulation is that it does not make any assumption about the discretization of the time axis. This is because the features $\vec\xi$ in Eq.~\eqref{eq:f windowed decomposition} are not directly dependent on the local time axis $\tau$ (see Figure~\ref{fig:features_illustration}), only the transform to extract them is. 
In other words, the specific discretization choices of the sliding time window $\eta(t,T)$ used in practice only affect how the features are extracted (and therefore possibly the accuracy of their extraction), but not their interpretation or physical meaning.
To give a concrete example, assume that the chosen features of a given continuous window $\zeta_p(t^*,\tau)$ can be exactly represented by a second order polynomial. 
The set of features $\vec\xi(t^*)$ is then given by the three polynomial coefficients $\vec\xi(t^*) = \left\{\xi_0(t^*), \xi_1(t^*), \xi_2(t^*)\right\}$ such that:
\begin{equation}\label{eq:polynomial example}
	\zeta_p(t^*, \tau) = \xi_0(t^*) + \xi_1(t^*) \tau + \xi_2(t^*) \tau^2, ~~\tau \in [0, T].
\end{equation}
If $\zeta_p(t^*, \tau)$ is then arbitrarily discretized on the time axis, the values of $\vec\xi(t^*)$ will remain unchanged. 
Nevertheless, their extraction accuracy from a discretized dataset may depend on the discretization itself. In the given example, at least three time samples would be needed to exactly determine the coefficient of a second degree polynomial interpolant.
This natural robustness to oversampling is especially interesting in the context of surrogate modeling of numerical dynamical systems, where the sampling frequency of the solver is often many times higher than the actual frequency content of the modeled signal responses.
It is also an important property for the mitigation of the over-reliance problem of classical NARX at high sampling rates (see Section~\ref{sec:NARX_limitations}).

Note that $\vec\zeta (t)$ in Eq.~\eqref{eq:information_vector} is essentially a continuous equivalent of $\vec\varphi (t)$ in Eq.~\eqref{eq:varphi}, and that we can write the time window in a discretized form as follows:
\begin{equation}\label{eq:discrete_memories}
	\eta(t, T) = \{ t, t-\delta t, \dots, t - n_t\delta t  \}
\end{equation}
where $n_t = \lceil T / \delta t \rceil$.
% The key difference to the classical ARX modeling introduced in Section~\ref{sec:arx_modeling} lies in the use of the features $\vec\xi(t)$ generated through the functions $\mathcal{K}_1, \dots, \mathcal{K}_{(M+1)}$, rather than in the original set of input/outputs $\{\vec x, \vec y\}$.
The key difference to the classical ARX modeling introduced in Section~\ref{sec:arx_modeling} lies in the use of the temporal features $\vec\xi(t)$ over the sliding window $\eta(t, T)$, generated through the functions $\mathcal{K}_1, \dots, \mathcal{K}_{(M+1)}$.
Classical ARX modeling uses the values of the inputs (or transformations thereof, as in \citet{lee_2018}) at specific time-steps of the inputs and outputs. This still allow classical ARX to account for the temporal ordering of the data, but misses the opportunity to take advantage of the smoothness and compressibility in time of the process being modeled.

Because our approach relies on functional features of the inputs and outputs \textit{along the time axis} within a sliding window (see Figure~\ref{fig:features_illustration}), rather than individual time steps, we refer to it as \emph{functional nonlinear autoregressive with exogenous inputs} ($\mathcal{F}$-NARX) modeling.
A direct advantage of $\mathcal{F}$-NARX and the formulation in Eqs.~\eqref{eq:continuous_prediction} and \eqref{eq:information_vector} is that all the existing ARX-fitting strategies introduced in Section~\ref{sec:arx_modeling} can be adapted with minimal modifications if the problem at hand has a discretized time axis. 
Because the overwhelming majority of numerical modeling of dynamic systems relies on numerical computations, which are inherently discrete, in the remainder of this work we will focus on discrete-time systems and how these applications can benefit from $\mathcal{F}$-NARX models.

We start by observing that many of the feature extraction methods for continuous-time signals introduced in Section~\ref{sec:feature_centric_view} admit a discrete counterpart.
Classical examples include the discrete version of the Fourier transform \citep{Sundararajan_2001}, or principal component analysis \citep{pearson_1901} as the discrete alternative to the Karhunen-Lo\`eve expansion. 
Other popular methods include auto-encoders \citep{rumelhart_1986}, wavelet transforms \citep{Olkkonen_2011} and Isomaps \citep{tenenbau_2000}.
These methods can efficiently represent the memory of each variable, improving the scalability of $\mathcal{F}$-NARX models with respect to the effective system memories. 

\subsection{$\mathcal{F}$-NARX model fitting and prediction}\label{sec:fnarx_fitting_and_prediction}
We recall from Section~\ref{sec:arx_modeling} that to fit a classical NARX model, for each observation $(i)$ from the experimental design, we need to construct the design matrix $\vec{\Phi}^{(i)} \in \mathbb{R}^{\widetilde{N} \times n}$ and the corresponding output vector $\vec{y}^{(i)} \in \mathbb{R}^{\widetilde{N}}$ as defined in Eq.~\eqref{eq:Phi_matrix}.
A similar procedure can be followed with the $\mathcal{F}$-NARX model. Given an observation $(i)$, we keep our definition of the vector $\vec{y}^{(i)}$ but we also define the feature matrix $\vec{\Xi}^{(i)} \in \mathbb{R}^{\widetilde{N} \times \tilde{n}}$ as an equivalent to the matrix $\vec{\Phi}^{(i)}$:
\begin{equation}\label{eq:feature_matrix}
	\vec{\Xi}^{(i)} = \begin{pmatrix}
		\vec{\xi}^{(i)}(t_0) \\
		\vec{\xi}^{(i)}(t_0+\delta t) \\
		\vdots \\
		\vec{\xi}^{(i)}((N-1)\delta t)
	\end{pmatrix}.
\end{equation}
In analogy to Eq.~\eqref{eq:large_Phi_matrix}, the regression task can then be performed on the input matrix $\vec{\Xi}^{\text{ED}}$ and the output vector $\vec{y}^{\text{ED}}$ comprising the data of the full experimental design:
\begin{equation}\label{eq:large_Xi_matrix}
	\vec{\Xi}_\text{ED} = \begin{pmatrix}
		\vec{\Xi}^{(1)} \\
		\vdots \\
		\vec{\Xi}^{(N_\text{ED})}
	\end{pmatrix}, 
	\quad
	\vec{y}_\text{ED} = \begin{pmatrix}
		\vec{y}^{(1)} \\ 
		\vdots \\
		\vec{y}^{(N_\text{ED})} 
	\end{pmatrix}.
\end{equation}

Ending up with this classical regression setting is useful, for example, when the duration of individual realizations is long, or the experimental design is large.
In these cases $\vec{\Xi}^{\text{ED}}$ becomes very large and efficient regression fitting using \eg\ ordinary least squares as shown in Eq.~\eqref{eq:ordinary_least_squares} can be limited by the available computing resources, memory in particular.
In such a scenario, a viable solution is to perform the regression on a subset of the sample pairs instead of using the full matrix $\vec{\Xi}_\text{ED}$ and vector $\vec{y}_\text{ED}$. This subsampling approach has been recently discussed and exploited in \citet{Schaers_2024} and will also be used in the applications Section~\ref{sec:applications}.

During the prediction phase, the $\mathcal{F}$-NARX approach is similar to classical NARX modeling, in that it uses its own past predictions to predict new output time steps:
\begin{equation}\label{eq:FNARX_prediction}
	\widehat{y}(t+\delta t) = \widehat{\mathcal{M}}(\widehat{\vec{\xi}}(t+\delta t)),
\end{equation}
where $\widehat{\vec{\xi}}(t)=\left\{\vec{\xi}_{x_1}(t), \cdots, \vec{\xi}_{x_M}(t), \widehat{\vec{\xi}}_{y}(t)\right\}$ is the feature equivalent to the vector $\widehat{\vec{\varphi}}(t)$ in Eq.~\eqref{eq:varphi_prediction}, built using the previously predicted outputs:
\begin{equation}\label{eq:FNARX_prediction_details}
	\widehat{\vec{\xi}}_{y}(t) = \mathcal{K}_y(\widehat{\vec{\zeta}}_y(t)) 
	\quad
	\text{with}
	\quad
	\widehat{\vec{\zeta}}_y(t)=\widehat{y}(\eta(t-\delta t, T-\delta t)).
\end{equation}
Since the features $\widehat{\vec{\xi}}_{y}(t)$ are extracted from the model predictions themselves, the prediction process must be initialized using the initial values $\widehat{y}(\eta(T - \delta t, T - \delta t))$.
This requirement is derived from classical NARX modeling, as described in Section~\ref{sec:arx_modeling}, where the first $n_y$ values must be specified. 
Typically, the model prediction is initialized from zero or any other sensible values. For validation purposes, the model can also be initialized with the true values if available, especially if the evolution of the system is sensitive to these initial values.

Note the additional feature extraction step $\mathcal{K}_j$, which has to be performed at each prediction time step and for each input $j$. 
If a very high forecast rate is required, the transform $\mathcal{K}_j$ should therefore be computationally efficient to not compromise the forecast performance of the final surrogate.
It is worth noting, however, that this additional cost can be compensated for during subsequent processing of the vector $\widehat{\vec{\xi}}(t)$.
Consider the classical NARX prediction as shown in Eq.~\eqref{eq:osa} and \eqref{eq:varphi_prediction}. The mapping $\mathcal{G}$ is typically computationally more expensive if $\widehat{\vec{\varphi}}(t)$ is large, which is \eg\ the case if the system memory and thus the model orders are large.
The feature extraction step can produce a $\widehat{\vec{\xi}}(t)$ that is considerably smaller than $\widehat{\vec{\varphi}}(t)$ in the classical NARX setting, thus speeding up the evaluation of $\mathcal{G}$.

% Similar to the classical NARX model, we must initialize the forecast with the first $n_t$ (see Eq.~\ref{eq:discrete_memories}) response time steps of the system. For validation purposes, we initialize the prediction in the subsequent application in Section~\ref{sec:applications} with the true response values.

% 

\section{Sparse $\mathcal{F}$-NARX modeling using principal component analysis and polynomial regression}\label{sec:sparse_fnarx}
After introducing the general $\mathcal{F}$-NARX modeling approach in Sections~\ref{sec:feature_centric_view}-\ref{sec:fnarx_fitting_and_prediction}, we now establish a concrete implementation for applications following in Section~\ref{sec:applications}. 
This implementation utilizes principal component analysis (PCA) for the calculation of the temporal features and it additionally uses polynomial basis functions to introduce nonlinearity in the ARX prediction.

\subsection{Principal component analysis}
Let us consider a discretized version of the information matrix $\vec{\zeta} \in \mathbb{R}^{\widetilde{N} \times n}$ with rows $\vec{\zeta}(t) \in \mathbb{R}^{n}$ as introduced in Eq.~\eqref{eq:information_vector} using discrete time windows as defined in Eq.~\eqref{eq:discrete_memories}. 
Our goal is to apply a transform $\mathcal{K}_i$ to each element $\vec{\zeta}_i \in \mathbb{R}^{\widetilde{N} \times n_i}$ of $\vec{\zeta}$ to obtain a discrete feature matrix $\vec{\Xi}_i \in \mathbb{R}^{\widetilde{N} \times \tilde{n}_i} = \mathcal{K}_i(\vec{\zeta}_i)$.
A well known tool to extract relevant features from auto-correlated discrete signals is given by principal component analysis (PCA) \citep{Jolliffe_2002}, which can be written as:
%Using principal component analysis (PCA), the transform involving $\mathcal{K}_i$ can be written as a linear transform:
\begin{equation}\label{eq:pca_mapping}
	\vec{\Xi}_i = \mathcal{K}_i^\text{PCA}(\vec{\zeta}_i) = \vec{\zeta}_i \vec{V}_i.
\end{equation}
The transformation matrix $\vec{V}_i$ is obtained by first standardizing $\vec{\zeta}_i$ to have zero mean and unit variance:
\begin{equation}
	\vec{Z}_i = \frac{\vec{\zeta}_i - \vec{\mu}_i}{\vec{\sigma}_i},
\end{equation}
where $\vec{\mu}_i \in \mathbb{R}^{n_{i}}$ and $\vec{\sigma}_i \in \mathbb{R}^{n_{i}}$ are the sample means and standard deviations of $\vec{\zeta}_i$, respectively. Subsequently, the covariance matrix $\vec{C}_i \in \mathbb{R}^{n_{i} \times n_{i}}$ is computed as:
\begin{equation}
	\vec{C}_i = \frac{1}{\widetilde{N} - 1} \vec{Z}_i^\top \vec{Z}_i,
\end{equation}
and an eigenvalue decomposition is performed to obtain its eigenvalues $\lambda_{ij} \in \mathbb{R}$ and eigenvectors $\vec{v}_{ij} \in \mathbb{R}^{n_i}$:
\begin{equation}
	\vec{C}_i \vec{v}_{ij} = \lambda_{ij} \vec{v}_{ij}.
\end{equation}
The matrix $\vec{V}_i$ then gathers the $\tilde{n}_i$ eigenvectors corresponding to the largest eigenvalues, in decreasing order of magnitude:
\begin{equation}\label{eq:projection_matrix}
	\vec{V}_i = \{ \vec{v}_{i1}, \vec{v}_{i2}, \ldots, \vec{v}_{i\tilde{n}_i} \}.
\end{equation}
In practical applications, $\vec{V}_i$ is usually truncated to a much smaller subset, corresponding to the largest $\tilde n_i$ principal components.
Consequently, the transform $\mathcal{K}_i^\text{PCA}$ can be parametrized in terms of the number of principal components $\tilde{n}_{i}$ where $1 \le \tilde{n}_{i} \le n_i$. 
A more intuitive approach to the parameterization is to adopt the \textit{explained variance} $\nu_i$, which corresponds to the fraction of the total variance of the signal that is reconstructed by the chosen truncated set, which can be computed as:
\begin{equation}\label{eq:explained_variance}
	\nu_i = \frac{\sum_{k=1}^{\tilde{n}_i} \lambda_{ik}}{\sum_{\ell=1}^{n_i} \lambda_{i\ell}}.
\end{equation}
By projecting each information vector $\vec{\zeta}_i(t)$ onto its corresponding PCA basis, we introduce a natural ranking in the features, as $\vec{v}_{i1}$ explains most of the variance of the $i$-th variable, followed by $\vec{v}_{i2}$, etc. 
This is different from classical ARX modeling, where such an order does not exist. For example, the shortest lags are not necessarily the most important ones, one of the reasons why the selection of lags is a well-known challenge in autoregressive modeling, as discussed in Section~\ref{sec:NARX_limitations}.
However, it should be noted that while $\vec{v}_{i1}$ explains most of the signal variance, this does not guarantee that it is also the most important mode for predicting the system response, which can also be governed by higher modes instead. 
Nevertheless, the parametrization of the $\mathcal{F}$-NARX model in terms of explained variance instead of discrete time lags can be considered an easier parametrization since a change in $\nu_i$ has more predictable consequences on the model performance compared to including or excluding individual lags.
A small $\nu_i$ leads to drastic dimensionality reduction. This simplifies the subsequent regression task but may discard principal components that are essential for accurate prediction, which can result in underfitting. 
In contrast, a large $\nu_i$ retains most of the signal variance and reduces information loss in the PCA step. 
However, it increases the dimensionality of the regression problem, which raises computational cost and the risk of overfitting to less relevant components.

\subsection{Polynomial $\mathcal{F}$-NARX model}\label{sec:polynomial_NARX}
Recall from Section~\ref{sec:arx_modeling} that NARX models are often formulated as linear-in-the-parameters models, leading to a linear regression problem as defined in Eq.~\eqref{eq:least_squares}, where a popular choice for the nonlinear mapping $\mathcal{G}$ is polynomials.
Polynomials have a long tradition in the use of NARX models and have proven to perform well in a variety of problems \citep{billings_2013}. 
In addition, they are easy to parameterize and computationally efficient.
In this section, we will explain how polynomials can be applied to build a polynomial $\mathcal{F}$-NARX model.

Given any discrete feature vector $\vec{\xi}(t) \in \mathbb{R}^{\tilde{n}}$ from the feature matrix $\vec{\Xi} \in \mathbb{R}^{\widetilde{N} \times \tilde{n}}$, we can construct the monomial $\mathcal{P}_{\vec{\alpha}}(\vec{\xi}(t))$ as follows:
\begin{equation}\label{eq:monomial}
	\mathcal{P}_{\vec{\alpha}}(\vec{\xi}(t)) = \prod_{i=1}^{\tilde{n}} \xi_i(t)^{\alpha_i},
\end{equation}
where $\xi_i(t)$ is the $i$-th component of $\vec{\xi}(t)$ and $\vec{\alpha} \in \mathbb{N}^{\tilde{n}}$ is an integer multi-index.
This allows us to approximate the output $y(t)$ as a weighted sum of monomials, where the weights $c_{\vec{\alpha}}$ are a set of real-valued coefficients:
\begin{equation}
	y(t) = \sum_{\vec{\alpha} \in \mathcal{A}} c_{\vec{\alpha}} \mathcal{P}_{\vec{\alpha}}(\vec{\xi}(t)).
\end{equation}
The multi-index set $\mathcal{A}$ is truncated to control the model complexity. In this study, we will use a truncation strategy that relies on three parameters: total polynomial degree $d$, maximum allowed interaction $r$, and hyperbolic truncation index $q$ as introduced in the context of polynomial chaos expansions by \citet{Blatman_2010}.
Consequently, the multi-index $\vec{\alpha} \in \mathcal{A}^{\tilde{n}, d, r, q}$ is constrained as follows:
\begin{equation}
	\mathcal{A}^{\tilde{n}, d, r, q} = \left\{ \vec{\alpha} \mid (||\vec{\alpha}||_0 \le r) \cap (||\vec{\alpha}||_q \le d) \right\},
\end{equation}
where $||\vec{\alpha}||_0 = \sum_{i=1}^{\tilde{n}} \mathbbm{1}_{\{\alpha_i > 0\}}$ and $||\vec{\alpha}||_q = \left(\sum_{i=1}^{\tilde{n}} \alpha_i^q \right)^{1/q}$ for $0 < q \le 1$.

Identifying the model coefficients $c_{\vec{\alpha}}$ can then be solved as an ordinary regression problem using \eg\ ordinary least squares as shown in Eq.~\eqref{eq:ordinary_least_squares} with regression matrix $\vec{\Psi} \in \mathbb{R}^{\widetilde{N} \times p}$ and output vector $\vec{y} \in \mathbb{R}^{\widetilde{N}}$:
\begin{equation}\label{eq:regression_matrix}
	\vec{\Psi} = \begin{pmatrix}
		\vec{\mathcal{P}}(t_0)\\
		\vdots\\
		\vec{\mathcal{P}}(t_\text{max})\\
	\end{pmatrix}, 
	\quad
	\vec{y} = \begin{pmatrix}
		y(t_0)\\
		\vdots\\
		y(t_\text{max})\\
	\end{pmatrix}.
\end{equation}
Here, we use $\vec{\mathcal{P}}(t)$ as shorthand notation for $\mathcal{P}_{\vec{\alpha}}(\vec{\xi}(t))$.

\subsection{Hybrid least-angle regression}\label{sec:LARS}

To solve the regression problem introduced in the previous section, we adopt least angle regression (LARS), a well-known sparse regression approach from the compressive sensing literature.
Least angle regression, first introduced in \citet{Efron_2004}, is a stepwise linear regression technique often used when the number of regressors is much larger than the number of samples in the regression problem.
Due to the use of lagged inputs and outputs, classical NARX models can fall into this category even when the dimensionality of the exogenous inputs is relatively low.
LARS has been successfully employed in the recent literature for the construction of sparse NARX models \citep{mai_2016, bhattacharyya_2020, Li_2021b}. 
Although $\mathcal{F}$-NARX modeling can reduce the dimensionality compared to classical NARX modeling, using a sparse solver can still be beneficial, especially when considering problems with high exogenous-input dimensionality, high polynomial degrees or high interaction orders (see Section~\ref{sec:polynomial_NARX}).

In this section, we describe how we modify the LARS algorithm, originally introduced by \citet{Efron_2004} and summarized in \ref{app:lars}, to fit an $\mathcal{F}$-NARX model tailored for surrogate modeling applications.
The proposed modification is particularly advantageous because surrogate modeling involves training the model on an experimental design that includes multiple realizations of the system being modeled (see Section~\ref{sec:arx_modeling}). 
The objective is to achieve optimal performance in long-term forecasts over a wide range of unseen inputs.
This goal differs with classical applications of the LARS algorithm in auto-regressive modeling (see, \eg \citet{zhang_2015}), which minimizes the one-step-ahead (OSA) prediction error rather than the forecast error. 
This choice is dictated by the observation that, in general, minimizing the OSA prediction error does not necessarily translate to stable and accurate forecast performance, especially when the forecast horizon is long.

In the literature, various methods have been developed to reduce the forecast error. These typically involve pruning, the process of selecting a single regressor to be eliminated at each iteration \citep{Piroddi_2003, Piroddi_2008, Piroddi_2010}, which can be computationally extremely expensive. Alternatively, gradient-based methods have been proposed, but they also incur high computational costs \citep{Farina_2009, Farina_2010}.
To obtain a $\mathcal{F}$-NARX model with good forecast performance while keeping model fitting costs manageable, we leverage the fact that LARS produces a sequence of coefficient vectors with increasing number of non-zero elements, which we refer to as the LARS path.

Let us denote as $\widehat{\vec{c}}^{(k)}$ one such coefficient vector from the LARS path.
For a coefficient vector $\widehat{\vec{c}}^{(k)}$, we compute the mean forecast error over the experimental design using the normalized mean squared error:
\begin{equation}\label{eq:lars_forecast_error}
	\varepsilon_y^{(i)}(\widehat{\vec{c}}^{(k)}) = \frac{1}{N_t} \frac{
		{\displaystyle \sum_{t=t_0}^{t_\text{max}}} \left[ y^{(i)}(t) - \widehat{y}^{(i)}(\widehat{\vec{c}}^{(k)}, t) \right]^2
	}{
		\text{Var}(\vec{y}^{(i)}) + \gamma
	},
\end{equation}
where $\vec{y}^{(i)}$ is the output of the $i$-th realization in the experimental design, $\text{Var}(\vec{y}^{(i)})$ its variance, $\widehat{\vec{y}}^{(i)}$ is the corresponding model forecast, and $N_t = (t_\text{max} - t_0) / \delta t + 1$ is the number of forecasted time steps. The variable $\gamma$ is a small positive regularization term added to the denominator to avoid division by zero if a signal is constant. 
It can also prevent giving excessively high weight to signals with very low variance compared to the rest of the signals in the experimental design.
The mean forecast error value of interest is then computed as:
\begin{equation}\label{eq:lars_mean_forecast_error}
	\overline{\varepsilon_y}(\widehat{\vec{c}}^{(k)}) = \frac{1}{N_\text{ED}} \sum_{i=1}^{N_\text{ED}} \varepsilon_y^{(i)}(\widehat{\vec{c}}^{(k)}),
\end{equation}
and we select the coefficient vector that minimizes this mean forecast error over all LARS iterations as our final model coefficients:
\begin{equation}
	\widehat{\vec{c}} = \arg\min_k \overline{\varepsilon_y}(\widehat{\vec{c}}^{(k)}).
\end{equation}
This approach not only leads to better prediction performance, but it also reduces the risk of overfitting, as the coefficient vector is selected using a metric which is different to the one-step-ahead prediction error minimized by the original LARS algorithm from \citet{Efron_2004}.

Evaluating the forecast error for all coefficient vectors produced by LARS can be computationally demanding, as the number of vectors can reach $\mathcal{O}(10^{2-3})$ for complex systems requiring many active regressors.
In these cases, not evaluating all coefficient vectors $\widehat{\vec{c}}^{(k)}$ can be the preferred approach. For example, sets with very few non-zero coefficients are unlikely to be sufficient to describe very complex systems.
The fitting costs can be further reduced by only evaluating every $k$-th LARS iteration instead of evaluating after every iteration.

It is worth noting the similarity of this approach to the work of \citet{Blatman_2011}, which used LARS with a cross-validation scheme to select the best set of regressors for time-independent problems. 
The approach presented above can therefore be seen as an adaptation of their work to time-dependent problems.

\section{Applications}\label{sec:applications}
In this section we present the application of $\mathcal{F}$-NARX modeling to two case studies, namely:
\begin{itemize}
	\item a tall building under wind loading in Section~\ref{sec:building wind}, to demonstrate the robustness of the algorithm to changes in parameters such as the size of the training set, the sampling rate of the problem, and the memory window length;
	\item a three-story steel frame under seismic loading in Section~\ref{sec:steel frame}, to demonstrate the suitability of $\mathcal{F}$-NARX to emulate the extreme behavior of complex dynamic system for, \eg\ reliability and fragility analysis. 
\end{itemize}

\subsection{Eight-story building under wind loading}\label{sec:building wind}

\begin{figure}[!ht]
	\centering
	\includegraphics[width=0.7\textwidth]{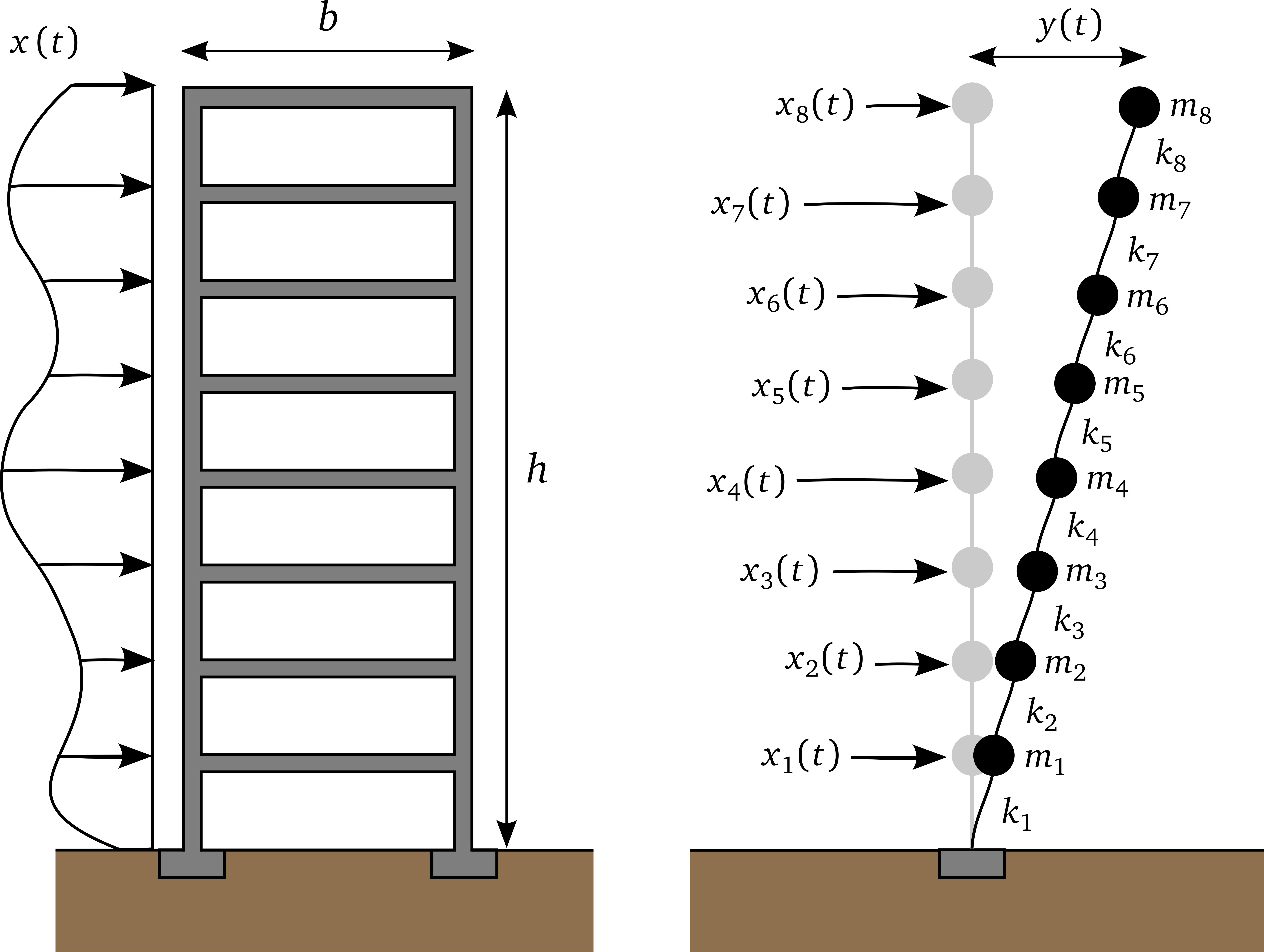}
	\caption{(Left) Sketch of the eight-story building model with horizontal wind loading $x(t)$. (Right) Corresponding lumped mass system with discretized wind loads $x_i(t)$(adapted from \citet{Jungho_2023}).}
	\label{fig:building}
\end{figure}

In our first case study, we consider the eight-story building displayed in Figure~\ref{fig:building}, which we adopted from \citet{Jungho_2023}.
The building is modeled as a linear elastic lumped mass system with the system parameters listed in Table~\ref{tab:building}.
The structure is subject to dynamic wind forces. The wind field generating these forces is modeled as a temporal random field, as described in \citet{Jungho_2023}. At a given time instant, the wind field consists of eight wind speed values corresponding to each of the eight building nodes shown in the right panel of Figure~\ref{fig:building}.
The parameters of the wind field model are mostly identical to those listed in \citet{Jungho_2023}.
The only exception is the basic wind speed $V_b$, which we model as a random variable following a normal distribution with a mean of $42.5$~m/s and a coefficient of variation of $33.3$~\%, in contrast to the deterministic value of $42.5$~m/s used in \citet{Jungho_2023}.
This choice is motivated by a need to increase the wind field complexity and variability to be closer to a realistic scenario.

\begin{table}[htb]
	\centering
	\caption{Eight-story building -- System parameters}\label{tab:building}
	\begin{tabular}{@{}lcc@{}}
		\toprule
		Parameter & Unit & Value \\ 
		\midrule
		Height $h$ & m & 32.3 \\
		Width $b$ & m & 36.6 \\
		Weight $m_1, \dots, m_8$ & kg & $9.66 \cdot 10^6$ \\ 
		Stiffness $k_1, \dots, k_8$ & N/m & $1.09 \cdot 10^9$ \\ 
		Damping ratio $c_1, \dots, c_8$ & - & 0.02 \\ 
		\bottomrule
	\end{tabular}
\end{table}

% \begin{figure}[!ht]
%     \centering
%     \includegraphics[width=\textwidth]{example_trace.png}
%     \caption{Example simulation: 30-second segment of the wind speed at the first, second and third node ($x_1(t)$, $x_2(t)$ and $x_3(t)$) and corresponding horizontal building top displacement $y(t)$.}
%     \label{fig:trace_illustration}
% \end{figure}

In this case study we assess the ability of $\mathcal{F}$-NARX to predict the horizontal top floor displacement $y(t)$ of the building. 
The exogenous model input is the wind field, comprising the eight wind speed time series $x_1(t), \dots, x_8(t)$.
We investigate the performance of $\mathcal{F}$-NARX for different configurations,  and assess its robustness to the time discretization of the problem.
We also compare the performance of $\mathcal{F}$-NARX to a classical NARX model. 
To perform this analyses, we use a set of $2{,}100$ system realizations, each with a duration of $10$~minutes and a sampling rate of $40$~Hz. 
The simulations were conducted using the MATLAB codes provided by the authors of \citet{Jungho_2023}.
A randomly selected set of $100$ of these simulations will serve as the experimental design, and the remaining $2{,}000$ simulations will be used as a hold-out validation dataset.

\subsubsection{Autoregressive model configurations}\label{sec:wind_building_model_configuration}
A total of 17 $\mathcal{F}$-NARX models, using the configurations listed in Table~\ref{tab:building_wind_model_structures}, were constructed.
The models are parameterized in terms of the explained variances $\nu_i$ and the model memories $T_i$. Note that the explained variances and memories were chosen homogeneously between the exogenous inputs and the autoregressive input. For simplicity, we will therefore refer to them as $\nu$ and $T$, omitting the subscript.
It is worth noting, however, that the building response may be influenced by both rapid fluctuations in wind speed and slower cumulative effects related to the damping of the building.
While a short memory may suffice to capture the wind effects, a longer memory might be necessary to adequately account for the damping effects.
Therefore, using a shorter memory for the exogenous wind related inputs could potentially enhance model performance or reduce the number of required model coefficients. Nevertheless, we observed that adopting a homogeneous memory approach yields good results and significantly simplifies the model parameterization.
We also vary the experimental design size $N_\text{ED}$ and the time discretization $\delta t$ of the problem. 
It is worth noting that the smaller experimental designs are a subset of the largest one and sampling rates other than $40$~Hz were not simulated but simply obtained by up-sampling or decimation of the $40$~Hz signals.

For Models~1-4, we keep $\nu$, $T$ and $\delta t$ fixed and only vary the experimental design size $N_\text{ED}$ from $1-100$ simulations. 
Models~5-10 have fixed $N_\text{ED}$, $T$ and $\delta t$, but we vary the explained variance $\nu$ from $0.85-0.995$.
For Models 11-13, $N_\text{ED}$, $\nu$ and $\delta t$ are fixed, while the model memory $T$ is varied from $1.0-4.0$~s.
Finally, for Models~14-17, we fix $N_\text{ED}$, $\nu$ and $T$ and vary the time discretization $\delta t$ from $0.1$~s down to $0.006$~s.

For each of the $17$ $\mathcal{F}$-NARX models we followed an adaptive approach to select the polynomial basis functions. 
We tested polynomial degrees ($d$) from $1$ to $3$, interaction orders ($r$) from $1$ to $3$, and q-norms of $0.7$, $0.85$, and $1.0$. 
The polynomial coefficients were calculated using hybrid LARS as described in Section~\ref{sec:LARS}.
We ran the LARS algorithm for a maximum of $200$ iterations and we evaluated the forecast performance on the experimental design for every $10$-th LARS iteration to speed up the model fitting process. 
To reduce computational costs, we used only a subset of at most $120{,}000$ rows from the design matrix to construct the surrogates, as described in Section~\ref{sec:fnarx_fitting_and_prediction}. This maximum was exhausted for all experimental design sizes except $N_\text{ED}=1$, which comprises only a total of about $24{,}000$ rows.
For all 17 models, the adaptive approach selected a setting of $d=1$, $r=1$ and $q=1$. Consequently, the algorithm identified a linear model to best describe the building response.
Please note that the algorithm is not limited to linear models, (see, e.g. the application discussed in Section~\ref{sec:steel frame}), but it automatic identified a linear model in this application as the best performing configuration.

In addition to the $\mathcal{F}$-NARX models, we also built a classical NARX model for comparison.
For this reference NARX model we used a model order of $15$ for both the exogenous inputs and the autoregressive component (see Eq.~\eqref{eq:varphi}), and we used the largest experimental design with $N_\text{ED}=100$ simulations for the model fitting.
A model order of $15$ was chosen to ensure a complexity comparable to that of the larger $\mathcal{F}$-NARX models. Moreover, it enables the use of nonlinear polynomials and the inclusion of interaction terms, which would become infeasible with higher model orders due to the rapidly growing number of polynomial terms.

To train the classical NARX model we followed the same basis adaptive approach with the hybrid LARS algorithm as used for the $\mathcal{F}$-NARX models.
For this reference model, the adaptive algorithm also selected a linear structure as the best-performing configuration, consistent with the findings for the $\mathcal{F}$-NARX models.

\begin{table}[]
	\centering
	\caption{Eight-story building -- $\mathcal{F}$-NARX model configurations}\label{tab:building_wind_model_structures}
	\begin{tabular}{@{}lcccc@{}}
		\toprule
		& \multicolumn{1}{l}{$N_\text{ED}$ {[}-{]}} & \multicolumn{1}{l}{$\nu$ {[}-{]}} & \multicolumn{1}{l}{$T$ {[}s{]}} & \multicolumn{1}{l}{$\delta t$ {[}s{]}} \\ \midrule
		Model 1  & 1                               & 0.950                                        & 1.0                                     & 0.025                                  \\
		Model 2  & 10                              & 0.950                                        & 1.0                                     & 0.025                                  \\
		Model 3  & 40                              & 0.950                                        & 1.0                                     & 0.025                                  \\
		Model 4  & 100                             & 0.950                                        & 1.0                                     & 0.025                                  \\
		Model 5  & 100                             & 0.850                                        & 1.0                                     & 0.025                                  \\
		Model 6  & 100                             & 0.900                                        & 1.0                                     & 0.025                                  \\
		Model 7  & 100                             & 0.930                                        & 1.0                                     & 0.025                                  \\
		Model 8  & 100                             & 0.970                                        & 1.0                                     & 0.025                                  \\
		Model 9  & 100                             & 0.990                                        & 1.0                                     & 0.025                                  \\
		Model 10  & 100                            & 0.995                                        & 1.0                                     & 0.025                                  \\
		Model 11 & 100                             & 0.950                                        & 0.5                                     & 0.025                                  \\
		Model 12 & 100                             & 0.950                                        & 2.0                                     & 0.025                                  \\
		Model 13 & 100                             & 0.950                                        & 4.0                                     & 0.025                                  \\
		Model 14 & 100                             & 0.950                                        & 1.0                                     & 0.100                                  \\
		Model 15 & 100                             & 0.950                                        & 1.0                                     & 0.050                                  \\
		Model 16 & 100                             & 0.950                                        & 1.0                                     & 0.0125                                  \\
		Model 17 & 100                             & 0.950                                        & 1.0                                     & 0.00625                                  \\ \bottomrule
	\end{tabular}
\end{table}

\subsubsection{Results}
\begin{figure}
	\centering
	\begin{subfigure}[b]{0.49\textwidth}
		\centering
		\includegraphics[width=\textwidth]{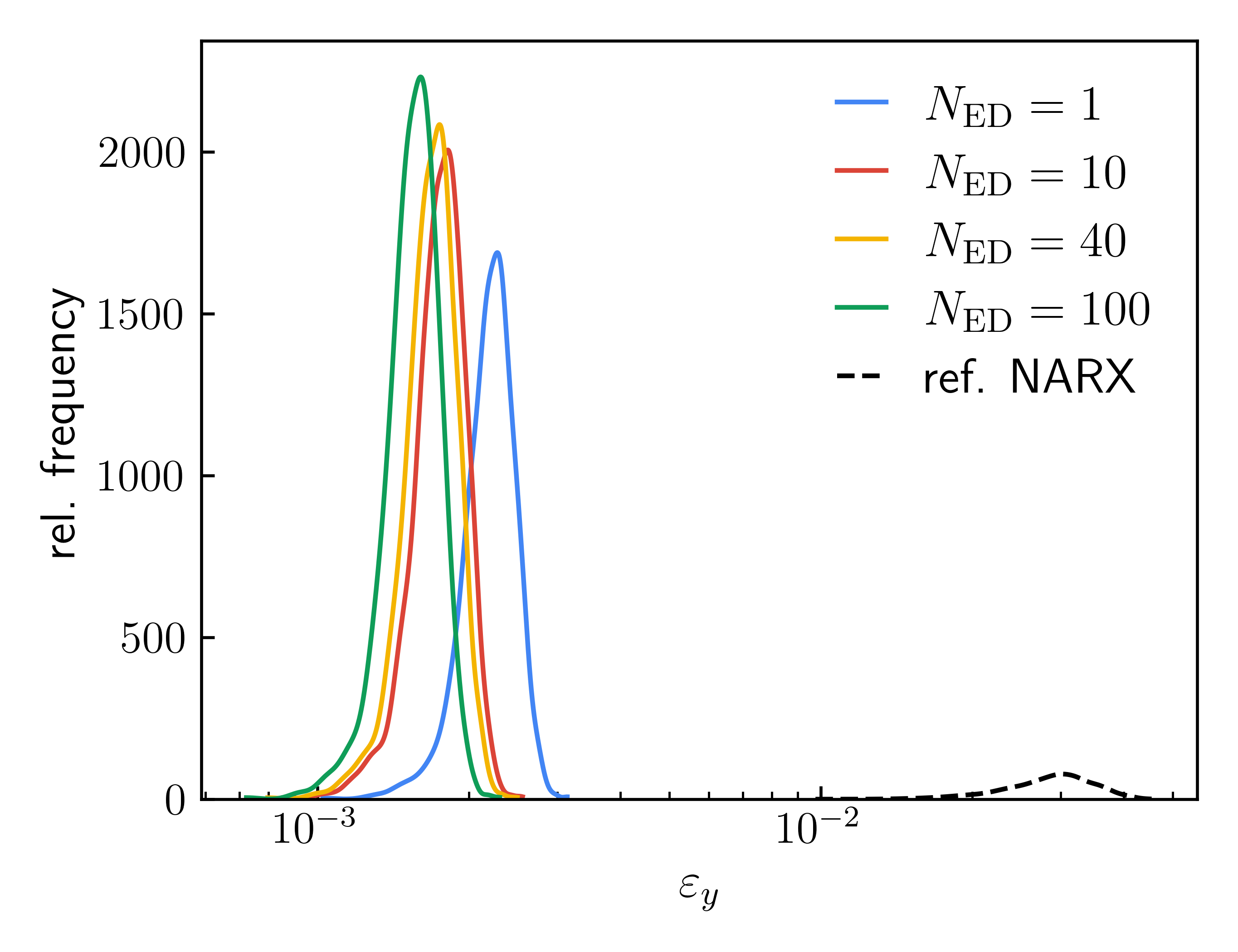}
		\caption{
			Prediction errors $\varepsilon_y$ as in Eq.~\eqref{eq:lars_forecast_error} on the validation dataset for different experiment design sizes $N_\text{ED}$. 
			The $\mathcal{F}$-NARX models correspond to Models~1-4 in Table~\ref{tab:building_wind_model_structures}.
		}
	\end{subfigure}
	\hfill 
	\begin{subfigure}[b]{0.49\textwidth} 
		\centering
		\includegraphics[width=\textwidth]{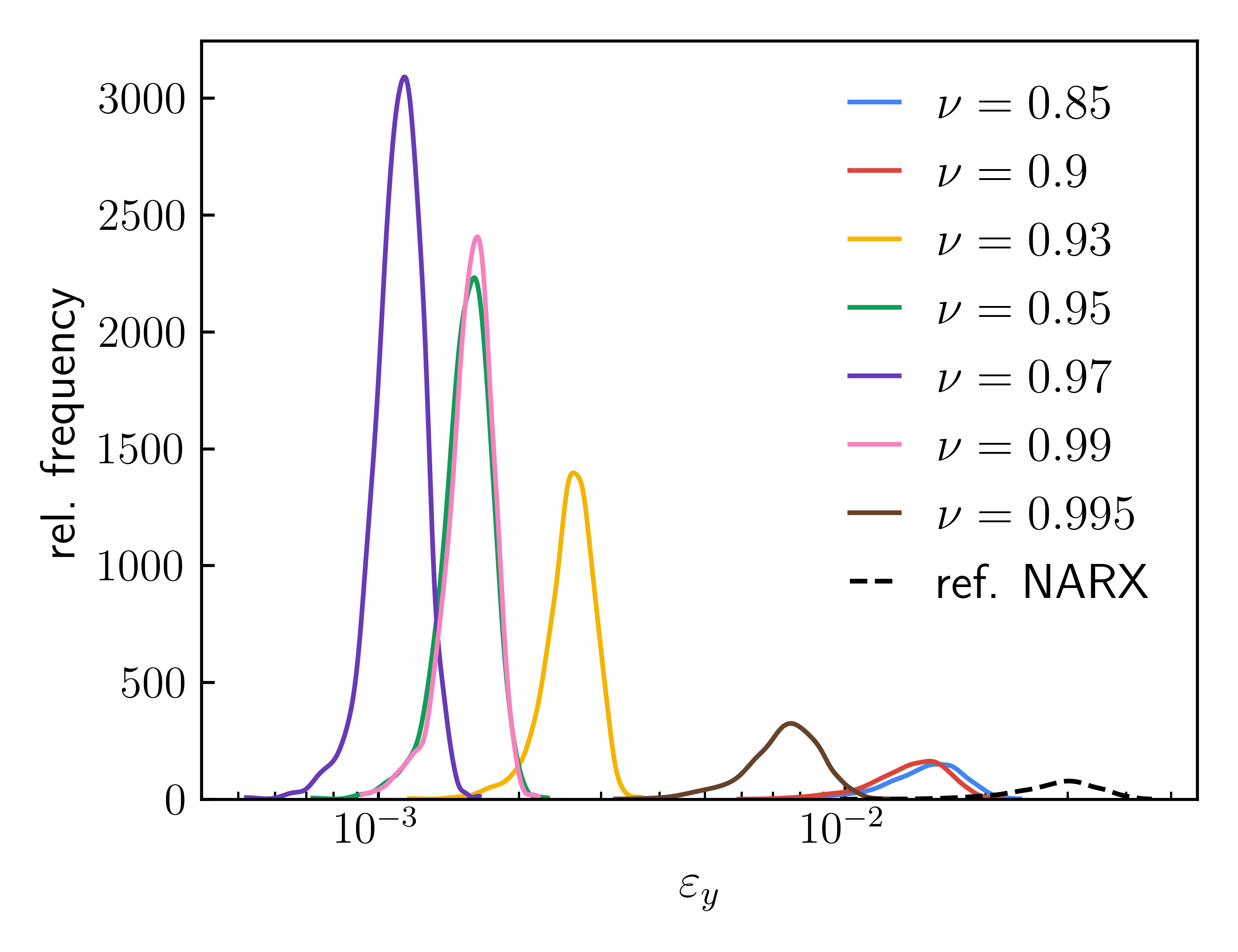}
		\caption{
			Prediction errors $\varepsilon_y$ as in Eq.~\eqref{eq:lars_forecast_error} on the validation dataset for different explained variances $\nu$. 
			The $\mathcal{F}$-NARX models correspond to Models~4-10 in Table~\ref{tab:building_wind_model_structures}.    
		}
	\end{subfigure}
	\vspace{1cm} 
	\begin{subfigure}[b]{0.49\textwidth}
		\centering
		\includegraphics[width=\textwidth]{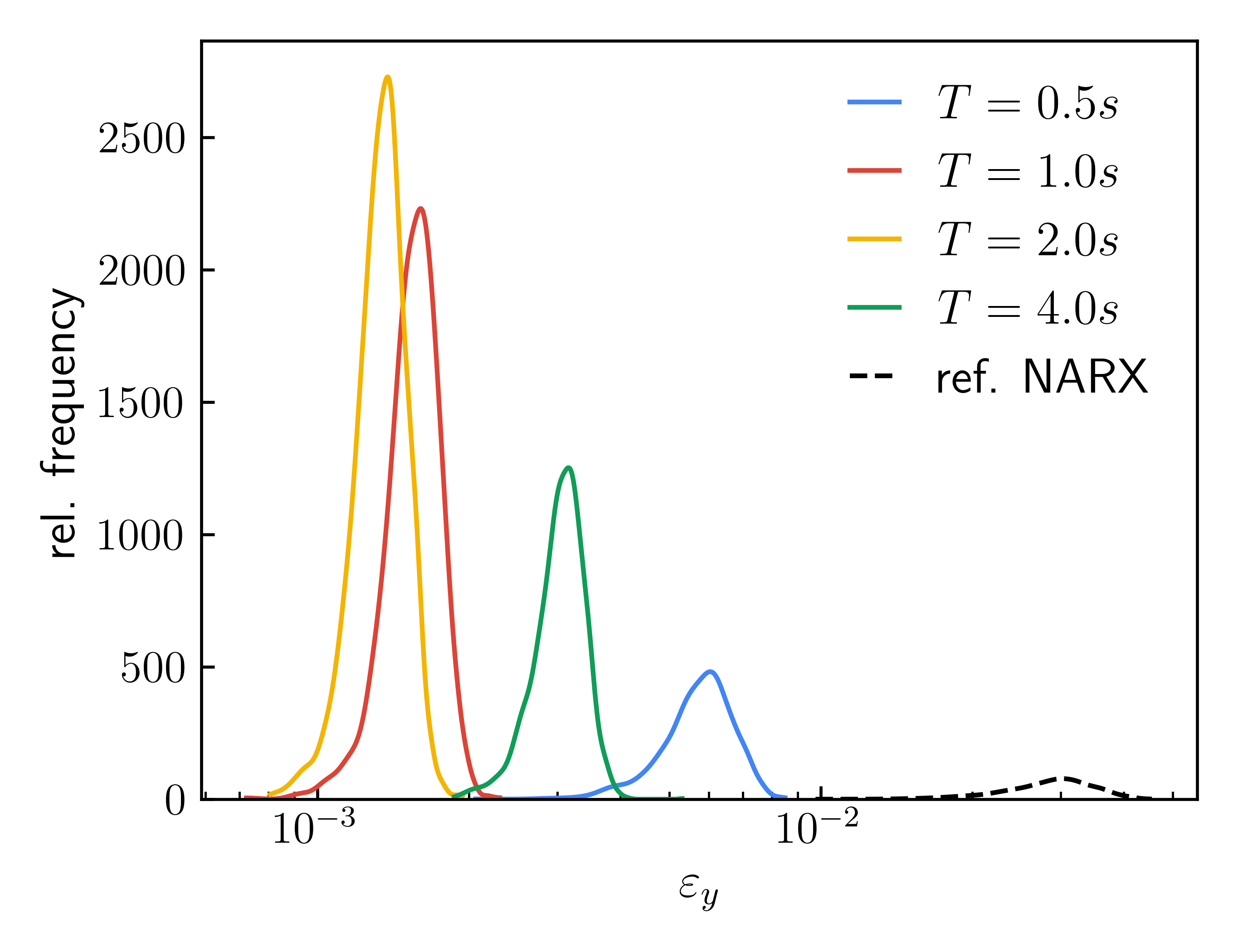}
		\caption{
			Prediction errors $\varepsilon_y$ as in Eq.~\eqref{eq:lars_forecast_error} on the validation dataset for different memories $T$. 
			The $\mathcal{F}$-NARX models correspond to the Models~4 and 11-13 in Table~\ref{tab:building_wind_model_structures}.
		}
	\end{subfigure}
	\hfill 
	\begin{subfigure}[b]{0.49\textwidth} 
		\centering
		\includegraphics[width=\textwidth]{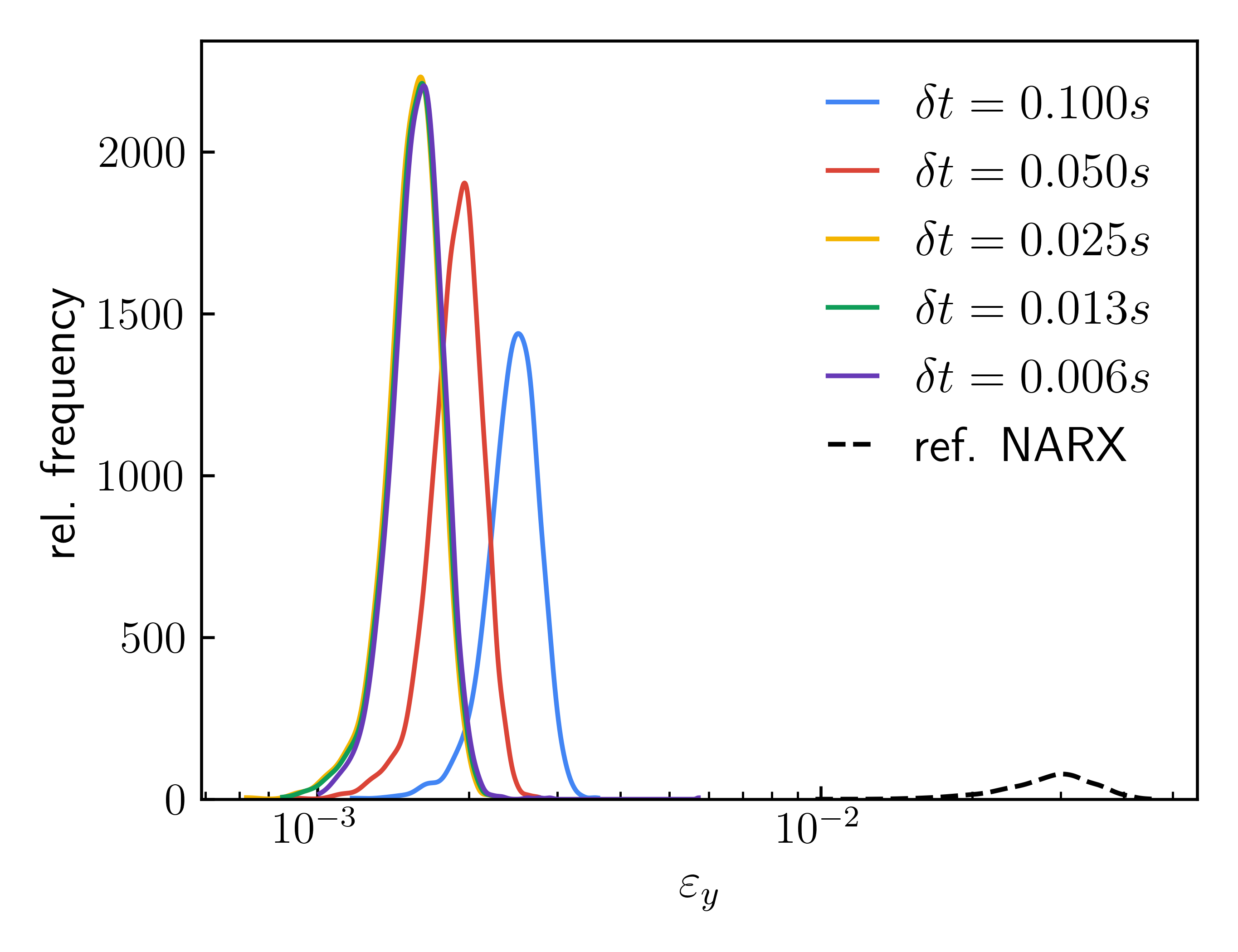}
		\caption{
			Prediction errors $\varepsilon_y$ as in Eq.~\eqref{eq:lars_forecast_error} on the validation dataset for different time increments $\delta t$. 
			The $\mathcal{F}$-NARX models correspond to Models~4 and 14-17 in Table~\ref{tab:building_wind_model_structures}.
		}
	\end{subfigure}
	\caption{
		Eight-story building results. 
		(a) Kernel density estimates of the distribution of the prediction errors $\varepsilon_y$ (see Eq.~\eqref{eq:lars_forecast_error}) on the validation dataset ($2{,}000$ curves) for different numbers of training simulations used to train the $\mathcal{F}$-NARX models (colored lines). The other configuration values were fixed to $\nu=0.95$ and $T = 1.0$~s and a time increment of $0.025$~s was used.
		The dashed black line shows the distribution of the prediction error for the classical NARX model (see Section~\ref{sec:wind_building_model_configuration}) for reference. Note that the abscissa is in log-scale.
		(b) Distribution of the prediction error for different values of the explained variance. All models were trained on $100$ simulations, the memory was fixed to $T = 1.0$~s and the data was sampled with $\delta t = 0.025$~s.
		(c) Prediction error for different model memories. All models were trained on $100$ simulations with the original time increment of $\delta t = 0.025$~s. The explained variance was set to $\nu=0.95$ for all models.
		(d) Prediction error for different resampling schemes (\ie, up-sampling or down-sampling). The original time increment is $\delta t = 0.025$~s. All models were trained on $100$ simulations and the other parameters were set to $\nu=0.95$ and $T = 1.0$~s.
	}
	\label{fig:building_convergence_results}
\end{figure}

\begin{figure}
	\centering
	\includegraphics[width=\textwidth]{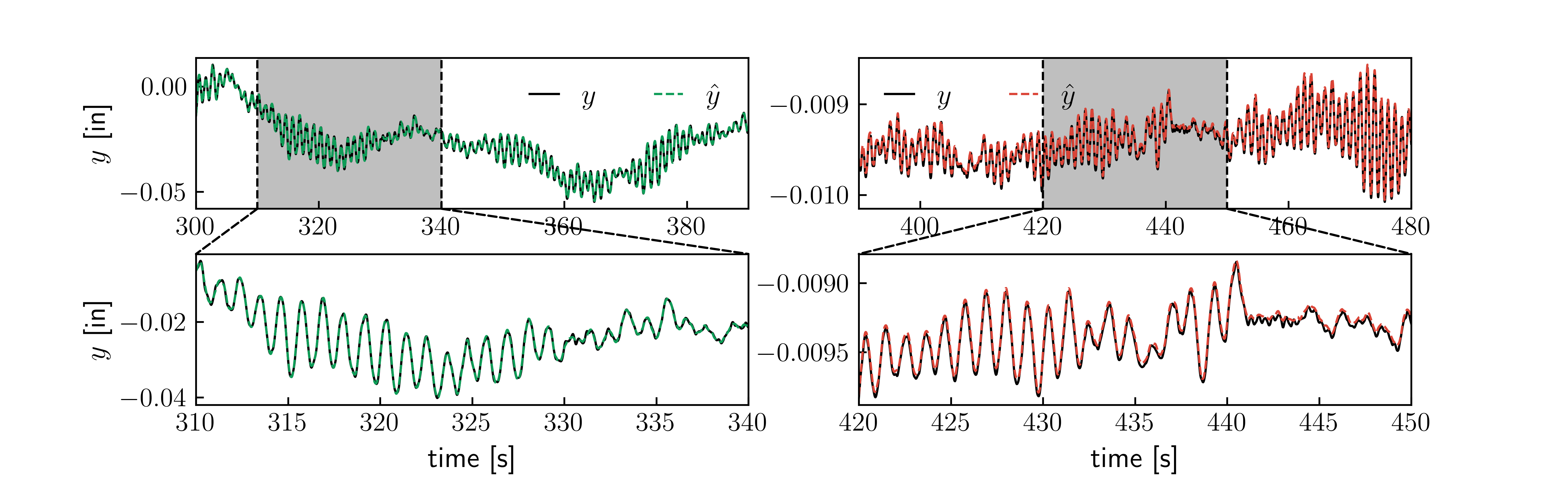}
	\caption{
		Eight-story building results. 
		(Left) Example traces showing the prediction of model~4 (see Table~\ref{tab:building_wind_model_structures}) with the lowest prediction error out of the $2{,}000$ validation curves. The colored line shows the model prediction while the black line shows the simulated output.
		Note that only a 90-second segment of the 10-minute-long simulation is shown.
		(Right) True vs. predicted trace with the highest prediction error out of the $2{,}000$ validation curves.
	}
	\label{fig:wind_building_traces}
\end{figure}

The performance of the $\mathcal{F}$-NARX models and the reference NARX model on the eight-story building under wind load are visualized in Figure~\ref{fig:building_convergence_results}. 
To quantify the goodness of the models we calculate the normalized mean-squared error as in Eq.~\eqref{eq:lars_forecast_error} with $\gamma=0$, and subsequently denoted as $\varepsilon_y$, for every trace in the validation dataset.
In Figure~\ref{fig:building_convergence_results}a, we show the kernel density estimate the distribution of all the $\varepsilon_y$ over the validation dataset for four different experimental design sizes $N_\text{ED}$. 
Consequently, the colored curves in this subplot correspond to the Models~1-4 from Table~\ref{tab:building_wind_model_structures}.
We also show in black the error distribution of the reference NARX model for comparison. 
Sections of the output traces corresponding to the best and worst prediction of Model~4 (out of $2{,}000$ validation curves) are shown Figure~\ref{fig:wind_building_traces} for illustration purposes.
In Figure~\ref{fig:building_convergence_results}a it can be seen that in general $\mathcal{F}$-NARX tends to result in more than one order of magnitude error reduction \wrt\ the reference NARX model.
It is also clear that $\mathcal{F}$-NARX models show a clear decreasing error trend with increasing experimental designs. 
Interestingly, even a single trace in the experimental design can yield relatively good approximation error, at the cost of a much larger trace-to-trace accuracy variability. 

Figure~\ref{fig:building_convergence_results}b shows the $\mathcal{F}$-NARX performance for different values of the explained variance $\nu$ (see Eq.~\eqref{eq:explained_variance}). These models correspond to Models~4-9 in Table~\ref{tab:building_wind_model_structures}.
It can be seen that a $\nu$ smaller or equal to $0.93$ results in a clear loss in model performance. Very good performance is achieved for $ 0.95 \le \nu \le 0.99$ and the model performs best for $\nu=0.97$, showing that performance can be compromised if $\nu$ is chosen too large.
This can be explained by the increased dimensionality in the regression step for large $\nu$, which can still lead to overfitting, as we will discuss in more detail later. 
For small $\nu$ the model cannot capture enough information to approximate the system response.

The results for different memories $T$ are given in Figure~\ref{fig:building_convergence_results}c. 
The models in this figure correspond to Model~4 and Models~11 to 13 in Table~\ref{tab:building_wind_model_structures}.
Similarly to the previous case, both too large and too small values of $T$ lead to reduced performance, with an overly small value being more detrimental than an excessively large one. 
The best performance is achieved for $T = 2.0$~s, which corresponds to approximately twice the period of the first mode of the building ($T_1 = 1.13$~s).  
This behavior is likely observed because a memory that is too short prevents the model from capturing long-term effects. In contrast, a memory that is too long can make it more difficult for the model to focus on the relevant information within the sliding window, potentially leading it to learn spurious or physically irrelevant patterns.

Finally, Figure~\ref{fig:building_convergence_results}d showcases the behavior of $\mathcal{F}$-NARX different sampling rates of the simulations, compared to the original sampling interval $\delta t = 0.025$~s.
These correspond to Models~4 and 14 to 17 in Table~\ref{tab:building_wind_model_structures}.
It is apparent that upsampling of the data does not affect the model performance. 
However, if the data is downsampled excessively, the performance starts decreasing. This is expected as high-frequency information in the data necessary to describe the system response can be lost. 

\begin{figure}
	\centering
	\begin{subfigure}[b]{0.49\textwidth}
		\centering
		\includegraphics[width=\textwidth]{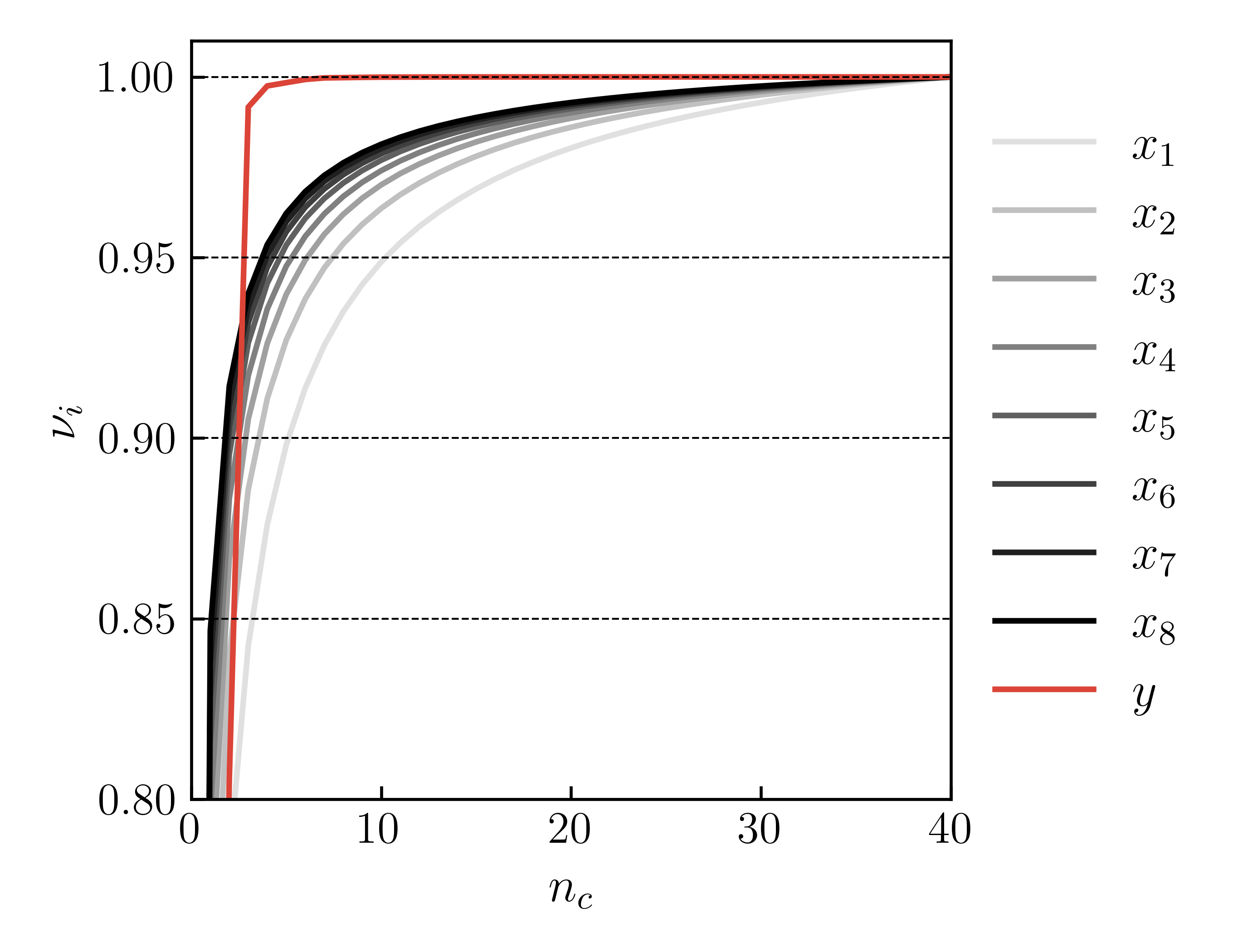}
		\caption{Number of principal components $n_\text{c}$ vs. explained variance $\nu_i$}
	\end{subfigure}
	\hfill 
	\begin{subfigure}[b]{0.49\textwidth} 
		\centering
		\includegraphics[width=\textwidth]{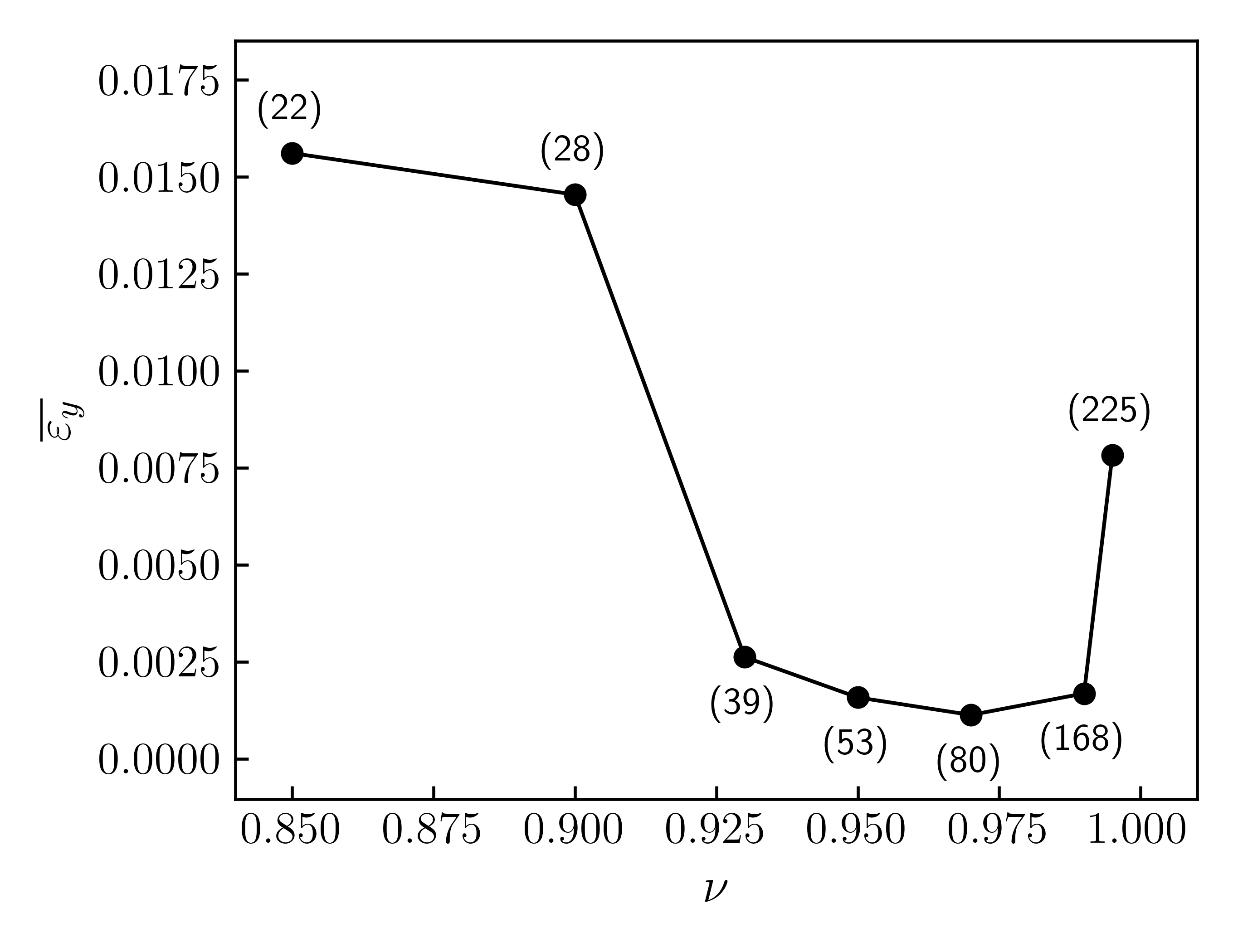}
		\caption{Mean prediction error $\overline{{\varepsilon}_y}$ for different values of $\nu$}
	\end{subfigure}
	\caption{
		Eight-story building results. 
		(Left) Convergence of the explained variances $\nu_i$ with respect to the number of principal components $n_{c}$ for each of the eight exogenous inputs $x_i$ and the output quantity $y$.
		(Right) Evolution of the mean prediction error $\overline{{\varepsilon}_y}$ (see Eq.~\eqref{eq:lars_mean_forecast_error}) for increasing values of $\nu$. The corresponding number of model coefficients is annotated in parentheses. 
	}
	\label{fig:building_helper_results}
\end{figure}

For a more thorough interpretation of the results from Figure~\ref{fig:building_convergence_results}, we plot the convergence of the explained variance $\nu_i$ for an increasing number of principal components $n_c$ and for each of the $\vec{\zeta}$ components (see Eq.~\eqref{eq:information_vector}) in Figure~\ref{fig:building_helper_results}a. Note that this plot corresponds to a memory of $T = 1.0$~s with a sampling rate of 40~Hz as \eg\ used for Model~4 in Table~\ref{tab:building_wind_model_structures}. 
We observe a fast initial increase in $\nu_i$ as $n_c$ gets larger. Consequently, most of the variance in the data can be explained by just a few components. 
The dimensionality during regression can thus be reduced significantly without much loss of information.
As expected, the explained variances for $y, x_1, \dots, x_8$ converge at different rates, with the model response $y$ converging the fastest. 
Since $\nu_i$ was chosen homogeneously, variables that show slow convergence contribute more to the total dimensionality because they require more components to reach the target explained variance. 
It can therefore be beneficial to cover less variance of variables with slow convergence if they are found to be less important to the model response.
An optimization of this trade-off could be performed through the dimensionality reduction surrogate modeling (DRSM) approach introduced in \citet{lataniotis_2020}, which performs a joint fit of the dimensionality reduction algorithm and surrogate model to improve the final surrogate accuracy. 

In Figure~\ref{fig:building_helper_results}b we show the evolution of the average $\varepsilon_y$ (average over the $2{,}000$ test simulations denoted as $\overline{\varepsilon_y}$) for increasing $\nu$.
For each data point, we annotate the corresponding number of regression coefficients, because it is expected to increase significantly with the explained variance $\nu$.
It can be observed that as $\nu$ increases, the model error first decreases with a large drop from $\nu=0.9$ to $\nu=0.93$. 
This can be explained by the fact that the first principal components explain a lot of variance and these components may also be the most important ones to explain the response dynamics without increasing the dimensionality of the problem much (see Figure~\ref{fig:building_helper_results}b).
As sufficient information is available to explain the response dynamics, the error plateaus until it increases again at about $\nu=0.99$. 
When $\nu$ approaches one, only comparatively little information is added by the extra input features, while the dimensionality increases rapidly and causes the model fitting to become underdetermined.

Note that the best value for $\nu$ for this problem lies within a relatively small window between $0.925$ and $0.975$.
However, choosing a value from this range appears natural due to the interpretability of this value. Alternatively, a plot such as shown in Figure~\ref{fig:building_helper_results}b can help in choosing an appropriate value.

\subsection{Three-story steel frame under seismic loading}\label{sec:steel frame}

\begin{figure}[!ht]
	\centering
	\includegraphics[width=0.7\textwidth]{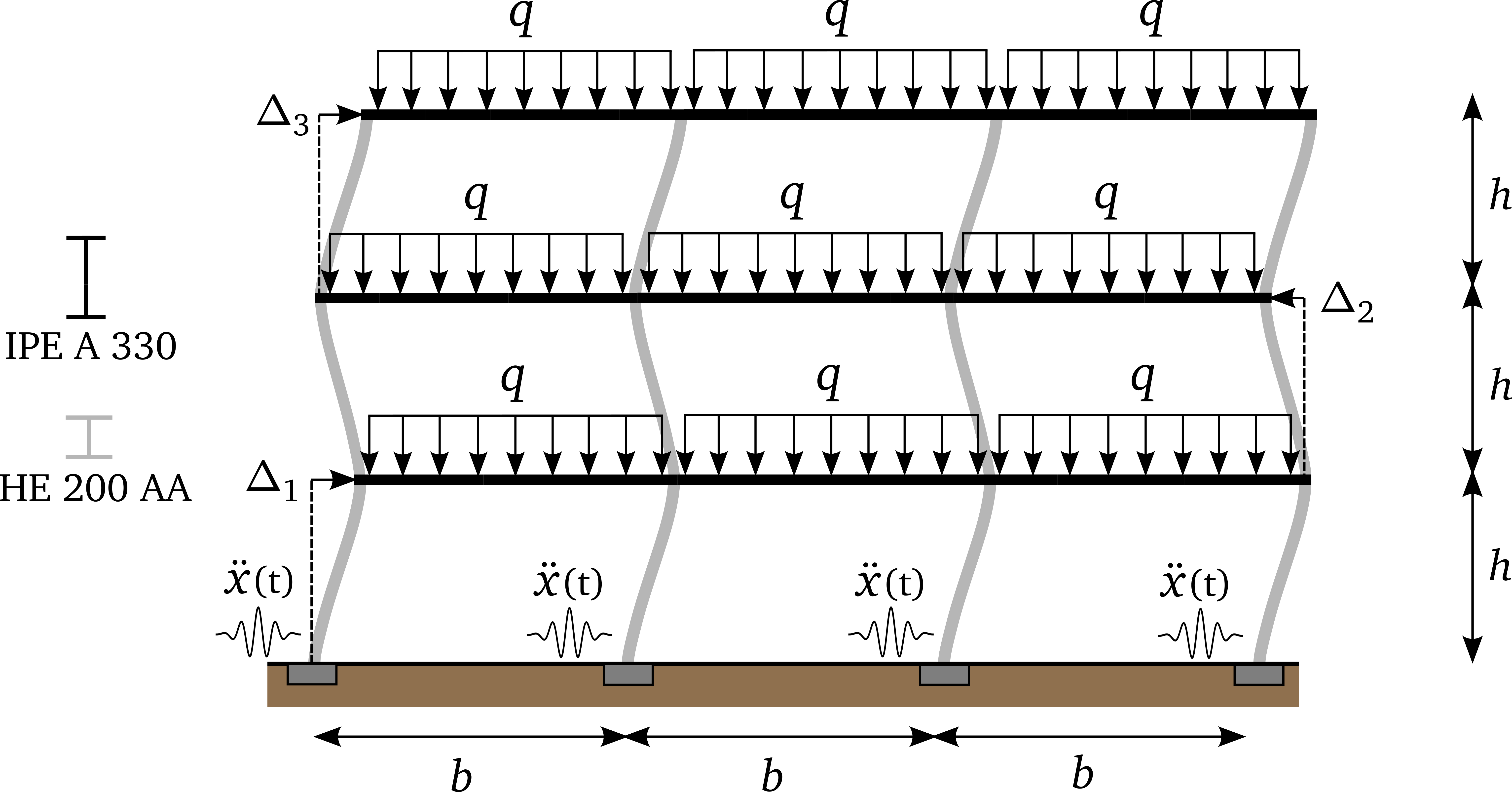}
	\caption{Three-story steel frame under ground motion excitation $\ddot{x}(t)$ and additional loading $q$. (Figure adapted from \citet{Zhu_2023})}
	\label{fig:steel_frame}
\end{figure}

In this application, we showcase the performance of $\mathcal{F}$-NARX on a realistic case study of a nonlinear three-story steel frame under seismic loading, which we adopted from \citet{Zhu_2023}. 
The frame structure is illustrated in Figure~\ref{fig:steel_frame}. Geometry parameters, basic material properties and the live load are listed in Table~\ref{tab:steel_frame}. 
For a detailed explanation of the model we refer to \citet{Zhu_2023}.

The quantities of interest are the three interstory drifts $\Delta_1(t), \Delta_2(t)$ and $\Delta_3(t)$ (see Figure~\ref{fig:steel_frame}). The exogenous inputs are the ground acceleration $\ddot{x}(t)$, ground velocity $\dot{x}(t)$ and the ground displacement $x(t)$ caused by seismic events. 
The ground acceleration data is taken from the Pacific Earthquake Engineering Research Center (PEER) database \citep{Power_2008} which gathers recordings of earthquake data near the earth's surface or on structure. 
In this case study we select earthquakes that leave the structure without significant permanent damage, as plastic deformation and damage accumulation would essentially require an infinite memory window. 
In turn, this behavior would make the problem unsuitable for a straightforward application of autoregressive modeling, which assumes finite system memory. 
Recent advancements on dynamic surrogate models, such as the recently introduced mNARX \citep{Schaers_2024}, could in principle address this type of behavior, but it remains outside the scope of this work. 

In total, we selected a rich set of $1{,}502$ acceleration recordings of real earthquakes sampled at $200$~Hz, covering near and far field, as well as low and high magnitude earthquakes.
The signal durations range from $10-300$~s with an average duration of about $75$~s.
The ground motion velocity and displacement were obtained by direct numerical integration of the acceleration data.
The simulations were performed using the open-source software OpenSees \citep{OpenSees}, and the resulting dataset was then randomly split into an experimental design comprising $100$ realizations and an out-of-sample test set of $1{,}402$ realizations for validation and performance assessment.

\begin{table}[htb]
	\centering
	\caption{Three-story steel frame -- Frame parameters}\label{tab:windmodel}
	\begin{tabular}{@{}lcc@{}}
		\toprule
		Parameter & Unit & Value \\ 
		\midrule
		Height $h$ & m & 3 \\
		Width $b$ & m & 5 \\
		Young's modulus $E$ & MPa & $2.05 \cdot 10^5$ \\
		Yield stress $f_y$ & MPa & 235 \\
		Live load $q$ & kN/m & 20 \\
		\bottomrule
	\end{tabular}
	\label{tab:steel_frame}
\end{table}

\subsubsection{Model configurations}
The configurations of the three final surrogates are shown in Table~\ref{tab:three_story_frame_model_config}.
These configurations were obtained by following a basis adaptive scheme as described in Section~\ref{sec:wind_building_model_configuration} using the hybrid LARS algorithm from Section~\ref{sec:LARS}.
We tested polynomial degrees ($d$) from $1$ to $3$, interaction orders ($r$) from $1$ to $3$, and q-norms of $0.7$, $0.85$, and $1.0$. 
We evaluated the forecast performance during LARS (Eq.~\eqref{eq:lars_forecast_error}) at every $10$-th iteration and stopped at a maximum of $500$ LARS iterations.
Further, we only used a subset of $100{,}000$ samples of the full design matrix to reduce the computational cost during the model fitting process.
The explained variances $\nu_i$ and memories $T_i$ were chosen identically for all three surrogates.
They were also chosen homogeneously between the exogenous inputs and the autoregressive input, so we will subsequently omit the subscript.

The memory length $T = 1.5$~s was chosen based on the period of the first mode of the structure, which is approximately $0.95$~s, indicating that a memory window in this range is necessary to capture the dominant dynamic behavior.  
To ensure sufficient coverage of these main effects, a slightly more conservative value of $T = 1.5$~s was selected.
This choice was also informed by the results from the first case study which found a memory of about $1.7$ times the first fundamental period to give the best results.

It is worth noting that the number of principal components required to achieve $\nu=0.95$ with  $T = 1.5$~s is considerably higher for the exogenous acceleration input $\ddot{x}$ than for the velocity $\dot{x}$ and displacement $x$. 
This is expected, since acceleration data carries higher frequencies than the corresponding velocities and displacements.
The best configuration for all the surrogates has a polynomial degree of three but only for $\Delta_2(t)$ interaction terms were included. 
Remarkably, the final surrogate for $\Delta_3(t)$ is significantly sparser (only 19 non-zero coefficients) compared to the other two surrogates.

Note that in this case study we do not make a comparison with a classic NARX model, as we were not able to produce stable long-term forecasts with it. In our tests, despite extensive trial-and-error, we could not eliminate frequently divergences in long-term predictions, making a meaningful performance comparison impossible. Similar instability issues with classical NARX models are well known in the literature \citep{Piroddi_2008, Farina_2009, Yu_2023}.

\begin{table}[]
	\centering
	\caption{Three-story steel frame -- Configurations of the three automatically selected $\mathcal{F}$-NARX models}\label{tab:three_story_frame_model_config}
	\begin{tabular}{@{}lccc@{}}
		\toprule
		Quantity of interest & $\Delta_1(t)$ & $\Delta_2(t)$ & $\Delta_3(t)$ \\ \midrule
		Memory ($T$)   & $1.5$~s        & $1.5$~s        & $1.5$~s        \\
		Explained variances ($\nu$)        & $0.95$      & $0.95$       & $0.95$       \\
		\# principal components ($n_{c,\ddot{x}}$,$n_{c,\dot{x}}$,$n_{c,x}$,$n_{c,\Delta}$) & $40$/$4$/$2$/$4$      & $40$/$4$/$2$/$3$       & $40$/$4$/$2$/$6$       \\
		Maximum polynomial degree ($d$)    & $3$          & $3$          & $3$          \\
		Maximum interaction order ($r$)        & $1$          & $2$          & $1$          \\
		Q-norm ($q$)                    & $1.0$       & $0.7$       & $1.0$       \\
		\# non-zero coefficients / \# coefficients & $146$/$150$        & $127$/$1{,}323$        & $19$/$156$        \\ \bottomrule
	\end{tabular}
\end{table}

\subsubsection{Results}
\begin{figure}%[ht]
	\centering
	\begin{subfigure}[b]{0.49\textwidth}
		\centering
		\includegraphics[width=\textwidth]{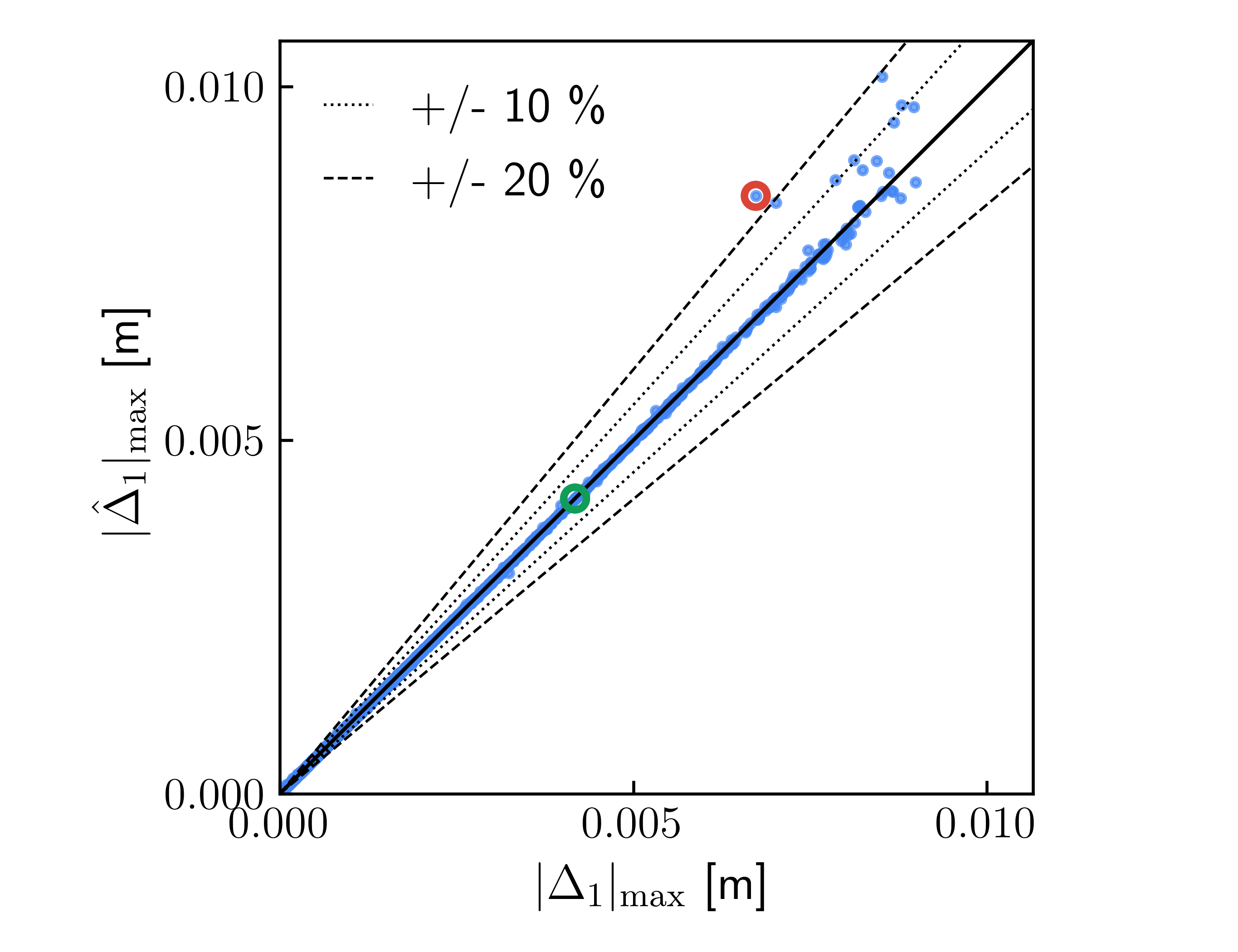}
		\caption{Predicted vs. true peak absolute 1$^\text{st}$ floor interstory drift}
	\end{subfigure}
	\hfill 
	\begin{subfigure}[b]{0.49\textwidth} 
		\centering
		\includegraphics[width=\textwidth]{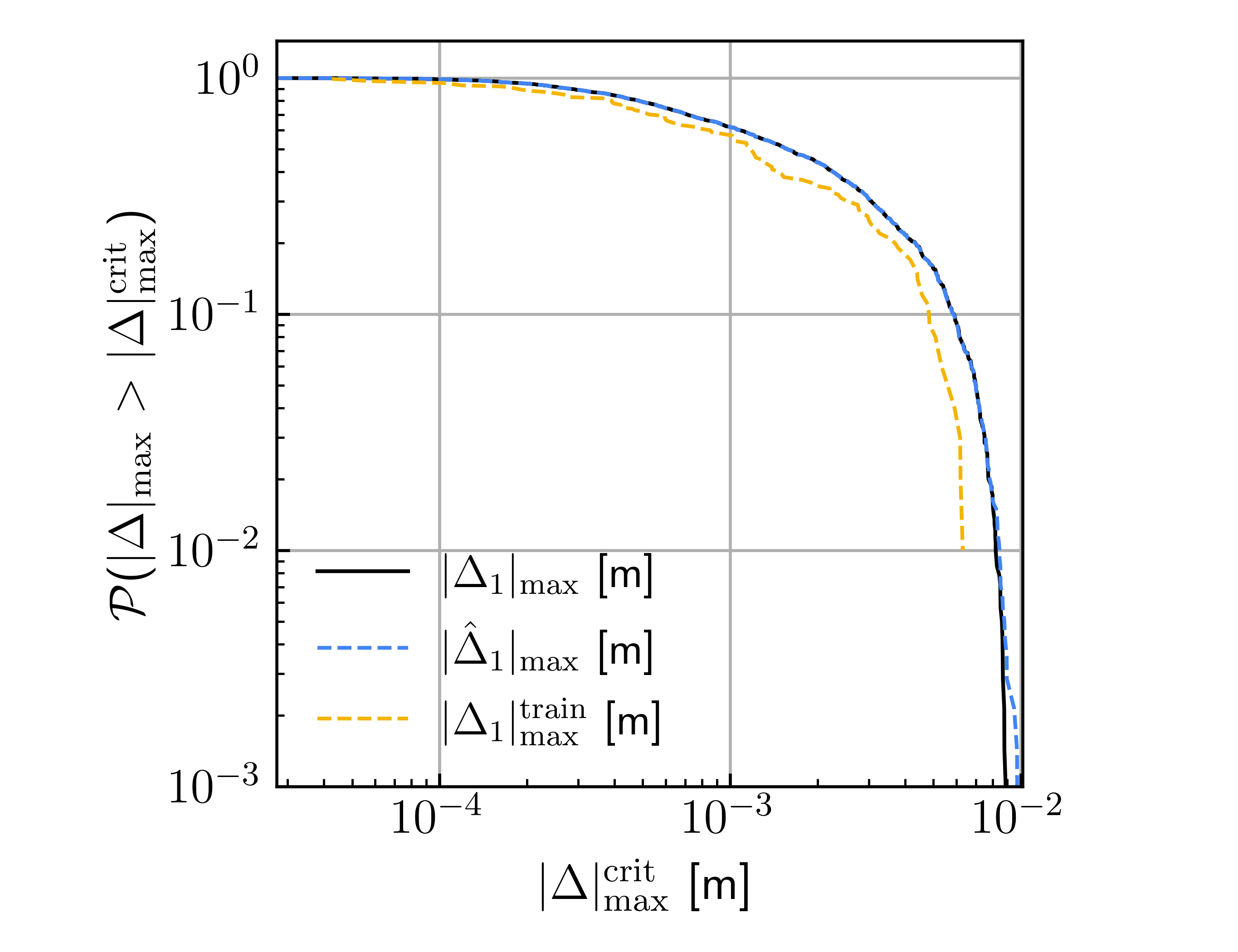}
		\caption{Survival curve for the peak absolute 1$^\text{st}$ floor interstory drift}
	\end{subfigure}
	
	\begin{subfigure}[b]{\textwidth} 
		\centering
		\includegraphics[width=\textwidth]{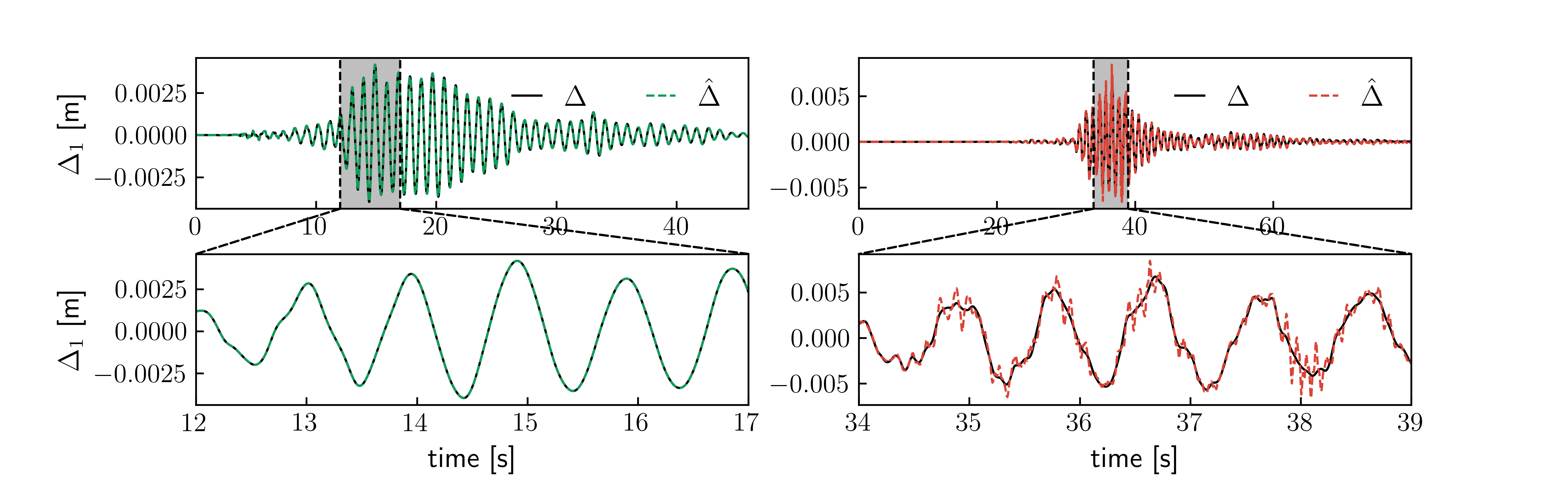}
		\caption{Example traces of the true (solid black lines) vs. predicted (dashed colored lines) 1$^\text{st}$ floor interstory drift}
	\end{subfigure}
	\caption{
		Three-story steel frame results. 
		(Top left) Scatter plot of the true vs. predicted peak absolute interstory drift of the 1$^\text{st}$ floor. The solid black line indicates a perfect prediction whereas the remaining lines show the $10-20$~\% error bounds.
		(Top right) Survival plot of the true and predicted peak interstory drifts. It shows the probability of $|\Delta|_\text{max}$ exceeding a given threshold value $|\Delta|_\text{max}^\text{crit}$. For reference the survival plot of the training dataset is shown in yellow.
		(Bottom) Example traces showing a good (green) and one of the worst (red) prediction. The green curve corresponds to a small relative error in the predicted absolute peak interstory drift. Analogously, the red curve has a high discrepancy in $|\Delta|_\text{max}$. Note that these traces correspond to the green and red circles in the top left panel.
	}
	\label{fig:earthquake_story1}
\end{figure}

\begin{figure}%[ht]
	\centering
	\begin{subfigure}[b]{0.49\textwidth}
		\centering
		\includegraphics[width=\textwidth]{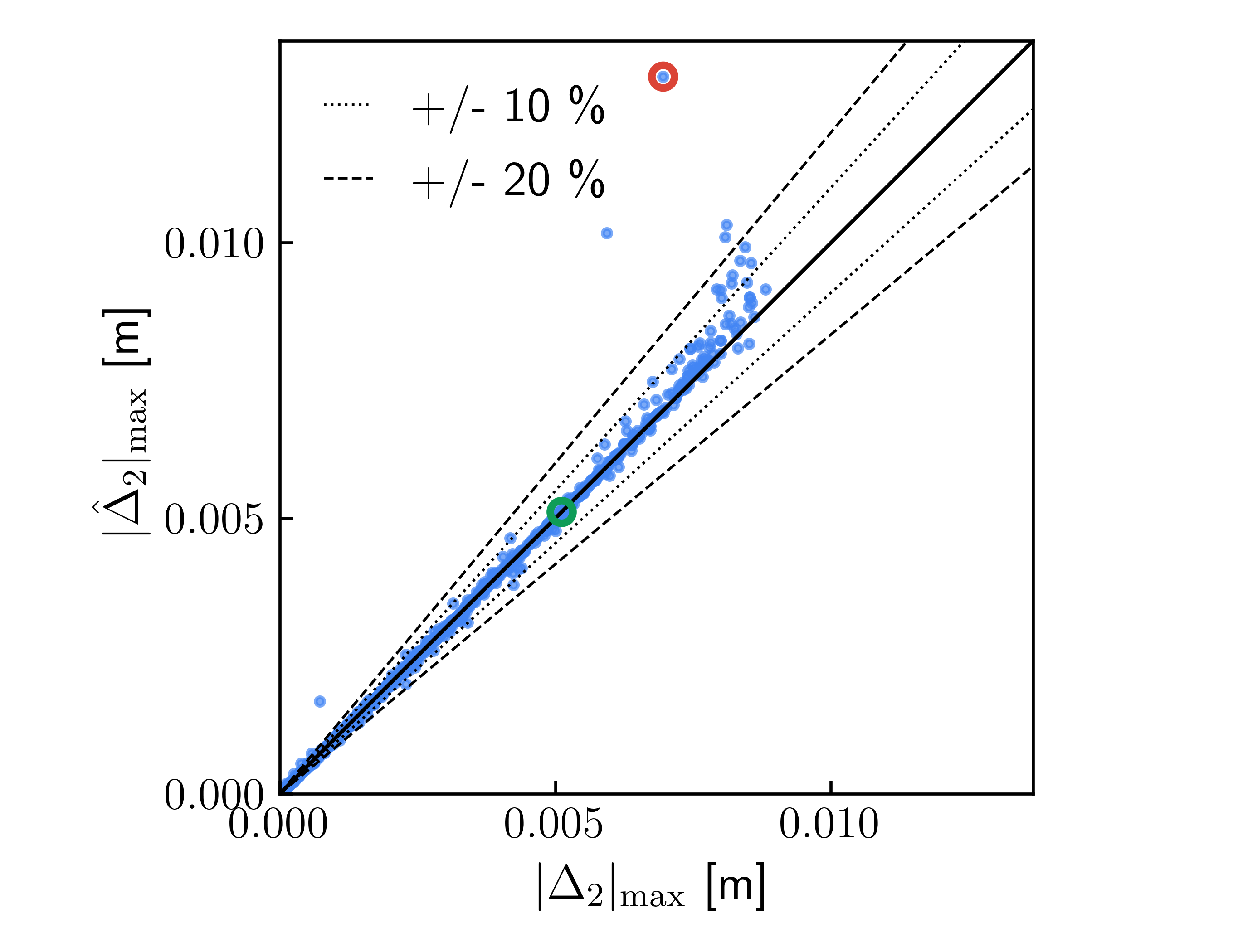}
		\caption{Predicted vs. true peak absolute 2$^\text{nd}$ floor interstory drift}
	\end{subfigure}
	\hfill 
	\begin{subfigure}[b]{0.49\textwidth} 
		\centering
		\includegraphics[width=\textwidth]{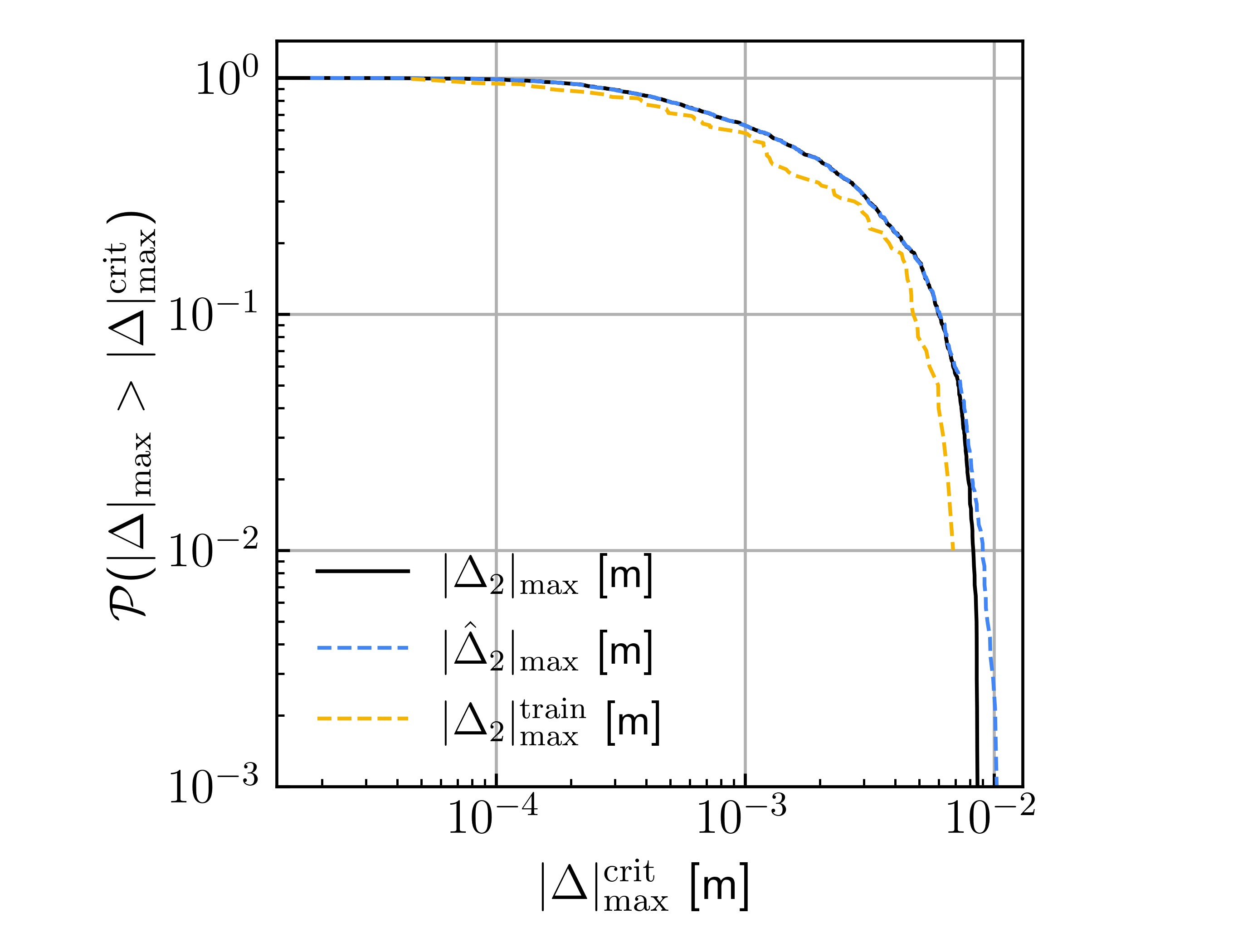}
		\caption{Survival curve for the peak absolute 2$^\text{nd}$ floor interstory drift}
	\end{subfigure}
	
	\begin{subfigure}[b]{\textwidth} 
		\centering
		\includegraphics[width=\textwidth]{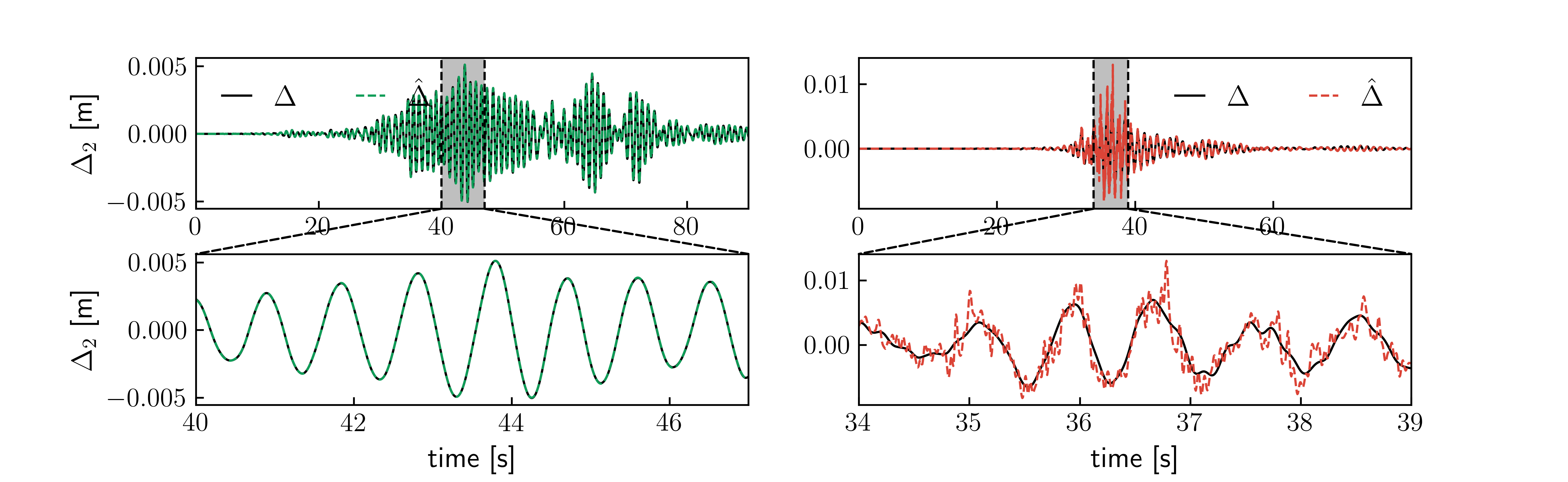}
		\caption{Example traces of the true (solid black lines) vs. predicted (dashed colored lines) 2$^\text{nd}$ floor interstory drift}
	\end{subfigure}
	\caption{
		Three-story steel frame results. 
		(Top left) Scatter plot of the true vs. predicted peak absolute interstory drift of the 2$^\text{nd}$ floor.
		(Top right) Survival plot of the true and predicted peak interstory drifts. 
		(Bottom) Example traces showing one of the best (green) and one of the worst (red) predictions. 
	}
	\label{fig:earthquake_story2}
\end{figure}

\begin{figure}%[ht]
	\centering
	\begin{subfigure}[b]{0.49\textwidth}
		\centering
		\includegraphics[width=\textwidth]{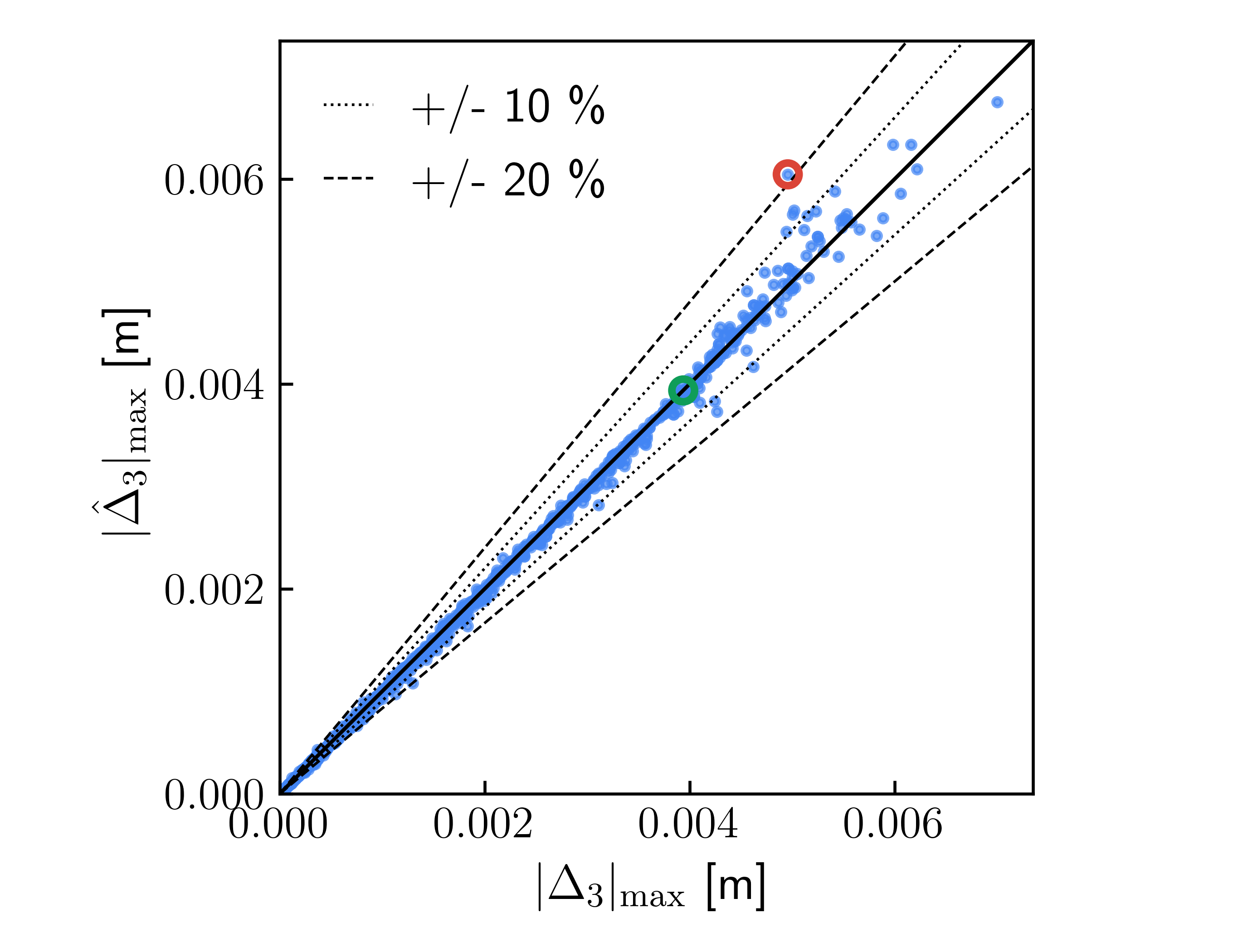}
		\caption{Predicted vs. true peak absolute 3$^\text{rd}$ floor interstory drift}
	\end{subfigure}
	\hfill 
	\begin{subfigure}[b]{0.49\textwidth} 
		\centering
		\includegraphics[width=\textwidth]{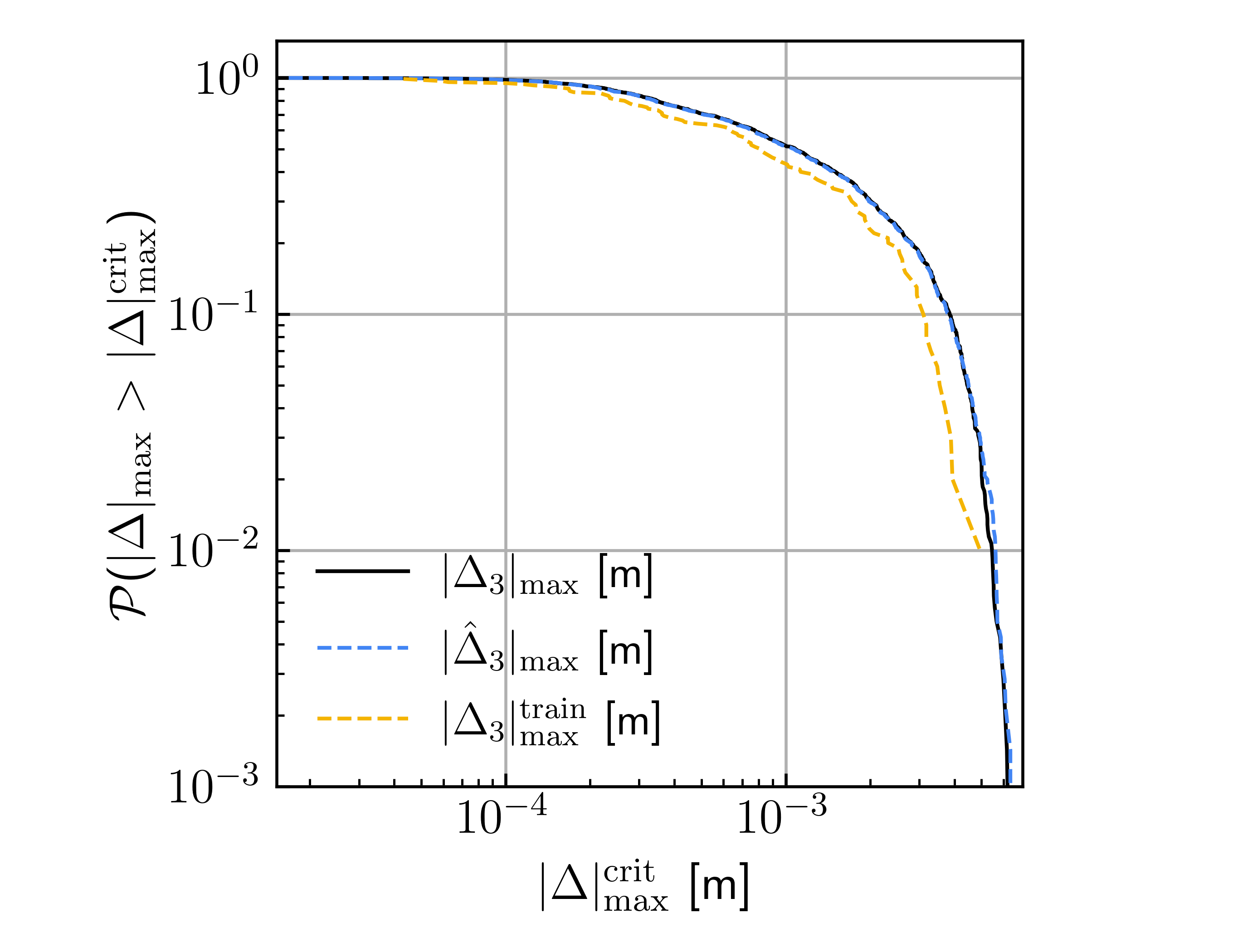}
		\caption{Survival curve for the peak absolute 3$^\text{rd}$ floor interstory drift}
	\end{subfigure}
	
	\begin{subfigure}[b]{\textwidth} 
		\centering
		\includegraphics[width=\textwidth]{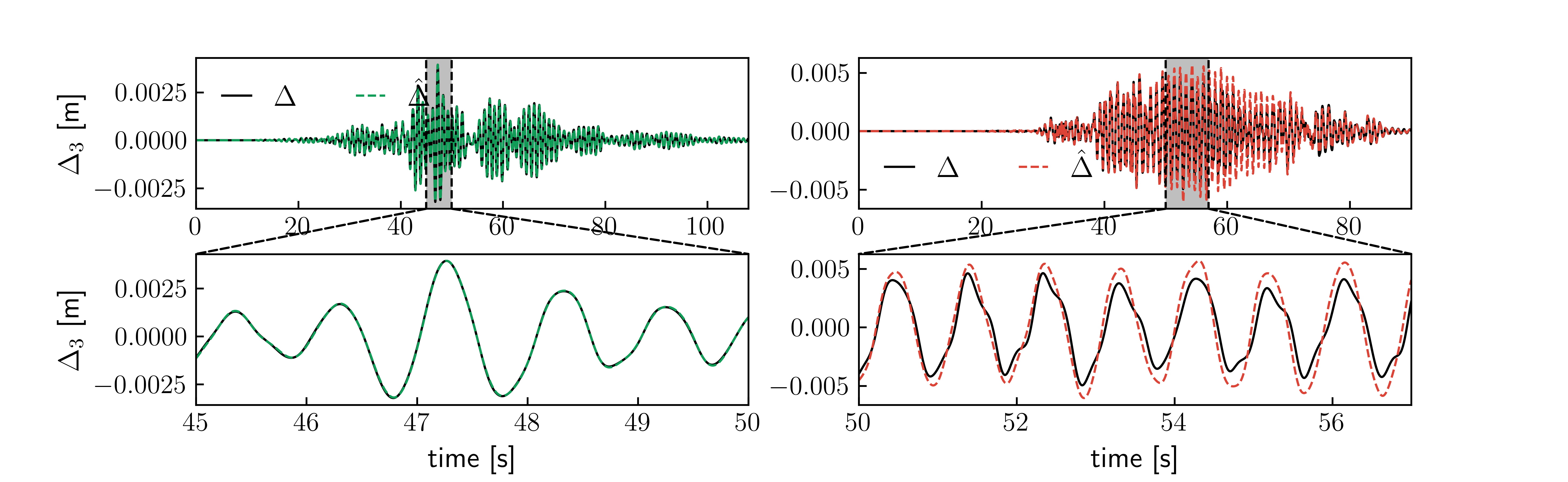}
		\caption{Example traces of the true (solid black lines) vs. predicted (dashed colored lines) 3$^\text{rd}$ floor interstory drift}
	\end{subfigure}
	\caption{
		Three-story steel frame results. 
		(Top left) Scatter plot of the true vs. predicted peak absolute interstory drift of the 3$^\text{rd}$ floor.
		(Top right) Survival plot of the true and predicted peak interstory drifts. 
		(Bottom) Example traces showing one of the best (green) and one of the worst (red) predictions. 
	}
	\label{fig:earthquake_story3}
\end{figure}

The results of the $\mathcal{F}$-NARX surrogates on the three-story steel frame case study are shown in Figures~\ref{fig:earthquake_story1}-\ref{fig:earthquake_story3}. 
Figure~\ref{fig:earthquake_story1}a compares the predictions of the $\mathcal{F}$-NARX model on the out-of-sample set of simulations for the 1$^\text{st}$~floor interstory drift, in terms of absolute peak drift, a quantity traditionally extremely difficult to approximate due to its highly non-linear and phase-sensitive nature.
It can be seen that the absolute peak interstory drifts of the predicted traces align well with the reference ones. For most simulations, the discrepancy is within the 10~\% error bounds, with only a handful of outliers with up to 30~\% error in the far right tail.
Figure~\ref{fig:earthquake_story1}b shows the corresponding survival plot with respect to a critical maximum drift $|\Delta|^\text{crit}_\text{max}$, calculated from the true and predicted drifts. 
The predicted curve matches the reference up to an exceedance probability of about $3 \cdot 10^{-3}$, with a mild deviation for lower probabilities.
For reference, the same curve computed from the $100$ training traces is provided, and as expected it clearly deviates from the reference already for relatively high exceedance probabilities because of the limited dataset. 
% This indicates that the surrogate can learn the system dynamics even on a not fully representative training set.
For visualization purposes, we also show two traces of two example simulations in Figure~\ref{fig:earthquake_story1}c, in which
the left and right traces have a low and high prediction error in $|\Delta|_\text{max}$, respectively.
The predicted trace with the low error aligns almost perfectly with the reference over the full simulation duration, while the other is generally stable over the full signal duration, but shows spurious higher frequency oscillations in the range of the extreme drift values.
To explain these oscillations, we must recall that the projection matrix used in the PCA step (see Eq.~\ref{eq:pca_mapping}) is estimated based on the $100$ traces used in the experimental design.
The out-of-sample validation traces may contain higher frequencies that are absent from the training data, leading the model to handle these frequencies incorrectly.
However, this also suggests that the model could be improved if such data becomes part of the experimental design.

Figures~\ref{fig:earthquake_story2}a-c show the same results for the 2$^\text{nd}$ floor interstory drift.
The results are mostly comparable to the ones for the 1$^\text{st}$ floor, albeit with an overall lower accuracy, especially in the high tail of the distribution.
The discrepancy in $|\Delta|_\text{max}$ is significantly higher, which is reflected in the higher dispersion of the points in Figures~\ref{fig:earthquake_story2}a and the earlier divergence between the true and predicted survival curves in Figures~\ref{fig:earthquake_story2}b.
The worse performance on the 2$^\text{nd}$ floor interstory drift may be due to more complex dynamics, which are also reflected in the higher polynomial degree and  interaction terms listed in Table~\ref{tab:three_story_frame_model_config}. 
% The surrogate predicting this floor is the only one having interaction terms (see Table~\ref{tab:three_story_frame_model_config}).

Finally, we show the results for the 3$^\text{rd}$ floor interstory drift in Figures~\ref{fig:earthquake_story3}a-c, which show an overall better agreement to the reference with respect to the previous two cases.
The fact that the surrogate predicting the 3$^\text{rd}$ floor has only $19$ non-zero model coefficients (see Table~\ref{tab:three_story_frame_model_config}) indicates that the dynamics of this floor are indeed somewhat less complex to predict than the others. 

\section{Discussion and Conclusion}\label{sec:discussion_and_conclusion}
In this paper, we present a novel approach to nonlinear autoregressive with exogenous inputs (NARX) modeling, called $\mathcal{F}$-NARX for \emph{functional} NARX. 
By following a continuous functional view point, rather than the well-established discrete-time approach, $\mathcal{F}$-NARX overcomes several limitations of traditional NARX models, making it particularly well-suited for surrogate modeling of dynamical physical systems, such as engineering structures, also in the presence of multiple exogenous inputs.
$\mathcal{F}$-NARX addresses the sensitivity of classical NARX models to time discretization, their tendency to over-rely on recent autoregressive inputs, and their inefficiency in handling systems with long memory.
These challenges are mitigated by modeling the system dynamics in a transformed space that captures key physical features from both the exogenous inputs and the autoregressive input (focus on functional features) rather than relying on original discrete time steps (focus on time discretization). 

To validate the methodology, we chose a combination of feature extraction using principal component analysis and sparse polynomial NARX modeling based on hybrid LARS, and applied it to two different case studies intended to showcase both the robustness of the method \wrt\ the limitations of classical NARX, and its performance in the presence of a complex computational model.

In the first case study, we examined the behavior of $\mathcal{F}$-NARX applied to an eight-story building subjected to wind loading.
The goal of this case study was to investigate the behavior of $\mathcal{F}$-NARX with respect to its configuration parameters.
We demonstrated that the model is straightforward to parameterize using interpretable configuration parameters, and at the same time that it is largely unaffected by changes in the sampling rate of the simulations, enabling it to handle highly oversampled signals effectively.
The main configuration parameter to choose, the model memory $T$, can in principle be optimized using cross-validation on an independent validation set. Alternatively, based on our findings, a value for $T$ in the range $[1.5T_0, 2T_0]$, where $T_0$ is the natural period corresponding to the lowest natural frequency of the system, serves as a good starting point.

In a challenging second case study, involving a nonlinear finite element model of a three-story steel frame under seismic loading, we demonstrated $\mathcal{F}$-NARX's capability to accurately model more complex dynamical systems. 
The $\mathcal{F}$-NARX models accurately predicted the interstory drifts of the building over extended periods, despite being trained on a small experimental design of only 100 simulations.
This forecast stability is largely due to the modeling of the system in the transformed feature space, which reduces the over-reliance on individual recent past output time steps that are highly correlated with the current output value.

The $\mathcal{F}$-NARX methodology, when deployed using PCA, polynomial basis functions with a basis adaptive scheme, and a sparse solver, proved to be a powerful surrogate for modeling complex dynamical systems typical of engineering simulation. Moreover, this implementation maintains low computational costs during model forecasting, making it practical for surrogate modeling.

For the first case study, the reference NARX model evaluates the validation dataset, comprising $2{,}000$ input samples with $24{,}000$ time steps each, in approximately $400$~s on a standard laptop. 
The $\mathcal{F}$-NARX model (Model~1) requires about $900$~s to evaluate the same dataset. 
This increase in evaluation time is due to the additional PCA projection step. 
It should be noted, however, that both the $\mathcal{F}$-NARX and classical NARX implementations are non-optimized in-house codes.

In the second case study, the $\mathcal{F}$-NARX model evaluates the $1{,}402$ validation traces in about $6$~minutes, which is approximately two orders of magnitude faster than the $\sim8$~hours required for the corresponding OpenSees simulations.

The more computationally expensive step is the model fitting, which uses basis adaptivity and a modified LARS algorithm (see Section~\ref{sec:LARS}). 
For the more complex second case study, the total cost of model fitting was about $6$~hours. 
The primary contributor to this cost is the repeated forecast error evaluation on the entire experimental design during the modified LARS procedure. 
Consequently, the time efficiency of this step depends heavily on the forecast efficiency of the surrogate model used. 
This cost would therefore be at most a factor of two lower for the classical NARX model. 
To reduce it significantly, either fewer coefficient sets along the LARS path can be evaluated, or only a subset of the experimental design can be used to estimate the forecast performance of the model.

Nevertheless, in realistic applications these training costs are largely offset by the gains in prediction performance, especially when dealing with complex computational models that often require minutes or even hours per run to execute.

Future research can explore the integration of other feature extraction algorithms and their synergy with different basis functions. 
Additionally, leveraging prior system knowledge to construct features relevant to the system response is a promising avenue. 
As shown by \citet{Schaers_2024}, incorporating expert knowledge into the modeling process can significantly improve the emulation of complex dynamical systems.

\section*{Acknowledgments}
This project is part of the \emph{HIghly advanced Probabilistic design and Enhanced Reliability methods
	for high-value, cost-efficient offshore WIND} (HIPERWIND) project and has received funding from the European Union's Horizon 2020 Research and Innovation Programme under Grant Agreement No. 101006689.

The authors express their sincere appreciation to Jungho Kim, Sang-ri Yi and Junho Song for their valuable contribution to the first case study by providing the corresponding numerical codes.

\appendix

\section{Least-angle regression}\label{app:lars}
The least-angle regression (LARS) algorithm, as presented in \citet{Efron_2004}, generates a sequence of coefficient vectors, often called the \emph{LARS path}, with each vector corresponding to one iteration. These vectors contain an increasing number of non-zero coefficients, as LARS adds one regressor per iteration, without removing any. 
The algorithm comprises the subsequent steps to select the regressors:
\begin{enumerate}
	\item \textbf{Initialization}: Let $\vec{c}^{(k)} \in \mathbb{R}^{p}$ be the coefficient vector at iteration $k$ and initialize it as a vector of all zeros: $\vec{c}^{(0)} = \vec{0}$.
	
	\item \textbf{Calculate residuals}: Compute the residual vector $\vec{y} - \vec{\Psi}_z\vec{c}^{(k)}$ where $\vec{\Psi}_z \in \mathbb{R}^{\widetilde{N} \times p}$ is the regression matrix as in Eq.~\eqref{eq:regression_matrix} standardized to have zero mean and unit variance, and $\vec{y} \in \mathbb{R}^{\widetilde{N}}$ is the corresponding output vector.
	
	\item \textbf{Find most correlated regressor}: Identify the regressor most correlated with the residuals and add it to the active set of regressors. We gather the indices of the active regressors in $\vec{\mathcal{A}}$.
	
	\item \textbf{Update coefficients}: Equally increase all the coefficients of the active regressors until one inactive regressor achieves the same correlation with the residuals as the current ones. 
	
	\item \textbf{Iterate}: Repeat steps 2 to 4 until all regressors are selected or a stopping criterion is met. In our work, we stop the algorithm when a given number of non-zero coefficients is reached. This number is a parameter of the algorithm.
\end{enumerate}

It is well known that the LARS-generated coefficients on the normalized and scaled data do not provide optimal prediction accuracy, and in their original work \citep{Efron_2004} introduced the idea of \textit{hybrid LARS-OLS}. 
Instead of directly using the coefficients estimated by LARS at any given point of its path, the ordinary least-squares solution can be computed on the original non-standardized data $\vec{\Psi}$ and regressors as follows:
\begin{equation}\label{eq:ols_minimization}
	\widehat{\vec{c}}^{(k)} = \mathop{\arg\min}_{\vec{c}^{(k)}_{\vec{\mathcal{A}}}} \| \vec{y} - \vec{\Psi} \vec{c}^{(k)} \|^2.
\end{equation}
Here, $\vec{c}^{(k)}_{\vec{\mathcal{A}}}$ are the coefficients corresponding to the regressors selected by LARS until iteration $k$, and their values are computed using ordinary least squares. 

\bibliography{bibliography}

\end{document}